\documentclass[%
 reprint,
 amsmath,amssymb,
 aps,
pra,
floatfix,
longbibliography,
]{revtex4-2}

\usepackage{graphicx}
\usepackage{dcolumn}
\usepackage{bm}
\usepackage{lipsum}
\usepackage{xcolor}
\usepackage{multirow}
\usepackage{svg}
\usepackage{appendix}
\usepackage{mathtools}
\usepackage{makecell}
\usepackage{amsmath}
\usepackage{float}
\usepackage{accents}

\DeclarePairedDelimiter\bra{\langle}{\rvert}
\DeclarePairedDelimiter\ket{\lvert}{\rangle}

\DeclarePairedDelimiterX\braket[2]{\langle}{\rangle}{#1\,\delimsize\vert\,\mathopen{}#2}
\newcommand\authormark[1]{\textsuperscript{#1}}

\begin{document}

\preprint{APS/123-QED}

\title{Photon pair generation {via down-conversion} in III-V semiconductor microrings: \\ modal dispersion and quasi-phase-matching}

\author{Samuel E. Fontaine,\authormark{1} Colin Vendromin,\authormark{1} Trevor J. Steiner,\authormark{2} Amirali Atrli,\authormark{1} \\Lillian Thiel,\authormark{3} Joshua Castro,\authormark{3} Galan Moody,\authormark{3} {John Bowers,\authormark{2,3}} Marco Liscidini,\authormark{4} and J. E. Sipe\authormark{1}}

\affiliation{\authormark{1}Department of Physics, University of Toronto, 60 St. George Street, Toronto, ON M5S 1A7, Canada\\ 
\authormark{2}Materials Department, University of California, Santa Barbara, CA 93106, USA\\
\authormark{3}Electrical and Computer Engineering Department, University of California, Santa Barbara, CA 93106, USA\\
\authormark{4}Dipartimento di Fisica, Università di Pavia, Via Bassi 6, 27100 Pavia, Italy} 

\email{samuel.fontaine@mail.utoronto.ca}

\date{\today}

\begin{abstract}
We explore how III-V semiconductor microring resonators can efficiently generate photon pairs and squeezed vacuum states {via spontaneous parametric down-conversion} by utilizing their built-in quasi phase matching and modal dispersion. We present an analytic expression for the biphoton wave function of photon pairs generated by weak pump pulses, and characterize the squeezed states that result under stronger pumping conditions. Our model includes loss, and captures the statistics of the scattered photons. A detailed sample calculation shows that for low pump powers conversion efficiencies of 10$^{-5}$, {corresponding to a rate of $39$ MHz for a pump power of 1 $\mu$W}, are attainable for rudimentary structures such as a simple microring coupled to a waveguide, in both the continuous wave and pulsed excitation regimes. Our results suggest that high levels of squeezing and pump depletion are attainable, {possibly leading to the deterministic generation of non-Gaussian states}.
\end{abstract}

\maketitle


\section{\label{sec:Intro}Introduction}

Spontaneous parametric down-conversion (SPDC) is routinely used to create photon pairs as well as squeezed states \cite{Gerry_Knight_2004}; the former are used for quantum key distribution \cite{Neumann2022experimental}, single heralded photon sources \cite{Herald}, and quantum communications \cite{Lu_2019}, while the latter are a resource for optical quantum computing \cite{Madsen2022QuantumCA} and quantum metrology \cite{Ligo}. In SPDC, pump photons interact with a material that has a second-order nonlinear susceptibility ($\chi^{(2)}$), and fission into pairs of lower energy signal and idler photons. The rates of these SPDC interactions are limited by the magnitude 
of $\chi^{(2)}$ and phase matching issues. {Yet,} in integrated {resonant} structures, such as microring resonators, {the light confinement} 
leads to large field intensities that enhance the effectiveness of the nonlinear susceptibility $\chi^{(2)}$, and for certain ring designs and certain choices of resonant modes the usual phase matching difficulties are alleviated \cite{Zhao:22,akin2024ingap}. 

For microrings fabricated out of III-V semiconductors, a ``built-in" quasi-phase-matching condition is present due to the difference between the crystal's natural reference frame and the local reference frame in which the fields propagate \cite{bib:Yang07,Yang:07_SPDC}. {In addition, III-V semi-conductors have very high values of $\chi^{(2)}$ compared to other materials commonly used for nonlinear interactions \cite{liu2023nonlinear}, leading to more efficient generation of quantum states. Also, recent advances in fabrication techniques have resulted in high-Q microring resonators useful for down-conversion from the visible to the telecommunication wavelengths have been developed \cite{Zhao:22,Thiel24}.} By adequately designing the microring resonator, the quasi-phase-matching condition can easily be satisfied, as the signal and idler effective refractive {indexes} can have values similar to that of the effective refractive index of the pump {due to} modal dispersion \cite{Zhao:22,akin2024ingap,Thiel24,Chang18,May:19}. 

{Experiments} where high rates of photon pairs were generated {have been reported} \cite{Zhao:22,akin2024ingap}, {and different approaches to arrive {at} rate equations or conversion efficiencies for microrings have been presented \cite{Yang:07_SPDC,bib:Guo16,bib:Brightness2019,Helt:12}. {In older work} losses {were neglected} \cite{Yang:07_SPDC,Helt:12,bib:Liscidini2012}, and {in more recent work, they have been treated by an approach based on Langevin equations} \cite{bib:Guo16,bib:Brightness2019}. In this work we extend these earlier calculations {by treating the SPDC interaction shown in \cite{bib:Liscidini2012} with the strategy for lossy structures presented in \cite{bib:Banic2022}. We also present extensive numerical simulations of realistic microring resonators from III-V semiconductors for down-conversion from the visible to the telecommunication range. Our calculation also extends beyond the pair regime, providing efficiencies for the squeezing regime.}

We present the generation efficiencies of photon pairs by SPDC in a lossy III-V semiconductor microring resonator point-coupled to a waveguide, where the waveguide is pumped by cw or pulsed excitation {in regimes where pump depletion can be neglected.} We account for loss by adding a point-coupled ``phantom" waveguide. Our approach uses  {an expansion in terms of asymptotic-in and asymptotic-out fields, which can be generalized to more complicated structures. This provides us with the statistics of not only the pairs of generated photons exiting the waveguide, but also the statistics of the pairs where both the photons are scattered, and those of the pairs where one photon is scattered and the other leaves the waveguide \cite{bib:Banic2022,bib:Liscidini2012}. {{Since} the information of the scattered photons is not lost, we retain a full characterization of the outgoing quantum state. {The results from our approach can also be used directly to treat more sophisticated problems involving high squeezing, pump-depletion, and non-Gaussian state generation \cite{Vendromin24}}.}}

{In the text we treat the scenario where the signal and idler photons generated are associated with different ring resonances (nondegenerate SPDC), but in Appendix B we present the corresponding expressions for the scenario where the signal and idler photons are associated with the same ring resonance (degenerate SPDC).}

{{For excitation by pump pulses we address two different regimes. In} the ``pair regime,"} the generated state consists {approximately} of {vacuum and a small probability amplitude of a pair} of photons that 
can be written {from} first-order perturbation theory as
\begin{equation}
    \ket{\psi}\approx \ket{\mathrm{vac}} + \beta \ket{{\mathrm{II}}}+...,
    \label{eq:PairRegime}
\end{equation}
where $\ket{\mathrm{II}}$ {is the ket {characterizing} {the} pair of photons,} and {$|\beta|^2$ can be {identified} as the probability {that a pair is generated. We find an analytic expression for the biphoton wave function (BWF) of the signal and idler photons.} In order for Eq. \eqref{eq:PairRegime} to be a valid approximation, we require that $|\beta|^2 \ll 1$. {We also} study realistic ring systems {and pumping scenarios} where {more than one pair is typically generated by each pump pulse, and} Eq. \eqref{eq:PairRegime} {no longer holds.} {We consider the regime where pump depletion can nonetheless still be neglected, and refer to this as} the ``lowest-order squeezing (LOS) regime." {Here} the ket for the generated photons is well approximated by}
\begin{equation}
    \ket{\psi} \approx  e^{\beta\int\mathrm{d}k_1\mathrm{d}k_2\varphi(k_1,k_2)a^\dagger(k_1)b^\dagger(k_2) - \mathrm{H.c.}}\ket{\mathrm{vac}},
    \label{eq:LOSR_Regime}
\end{equation}
{where $\varphi(k_1,k_2)$ is the {joint spectral amplitude,} and {the} $\beta$ {that appears here plays the role of a} ``squeezing parameter."} 

{The structure of this paper is as follows. In Sec. \ref{sec:Theory}, we begin by presenting the system and the theory to obtain the rate equation for cw excitation in the pair regime. We also present the BWF and the number of photons in the generated squeezed state for pulsed excitation for both the pair regime and the LOS regime.} In Sec. \ref{sec:NumResults}, we present a detailed sample calculation {of} the generation efficiencies {of} a realistic, simple structure: a microring coupled to a single physical waveguide, both fully encapsulated 
in silicon oxide (SiO$_2$) cladding. We show a number of structures for which {conversion} efficiencies are on the order of $10^{-5}$ at critical coupling for {cw} low pump powers, where the quasi-phase-matching and the dispersion properties of the system are taken into account, and show how the different coupling regimes compare with the one of critical coupling. We present various BWFs for the optimized structures for pulsed excitation for various pulse durations. The outgoing states have a range of different Schmidt numbers (\cite{bib:Houde2023,bib:Quesada2022,bib:Drago}). We also show that our optimized structures are potentially able to generate high amounts of squeezing 
and could possibly lead to pump depletion {-- which in turn could lead to deterministic generation of non-Gaussian states \cite{yanagimoto22} --} for relatively-low-energy pump pulses. Finally, in Sec. \ref{sec:Conclusion}, we include a brief conclusion, and discuss future work.

\section{\label{sec:Theory}Theory}
We consider a microring resonator close to a waveguide, and formally introduce an artificial, ``phantom" waveguide as well to model scattering losses \cite{bib:Liscidini2012,bib:Banic2022}.
\begin{figure}[ht]
    \centering
\includegraphics[width=0.40\textwidth]{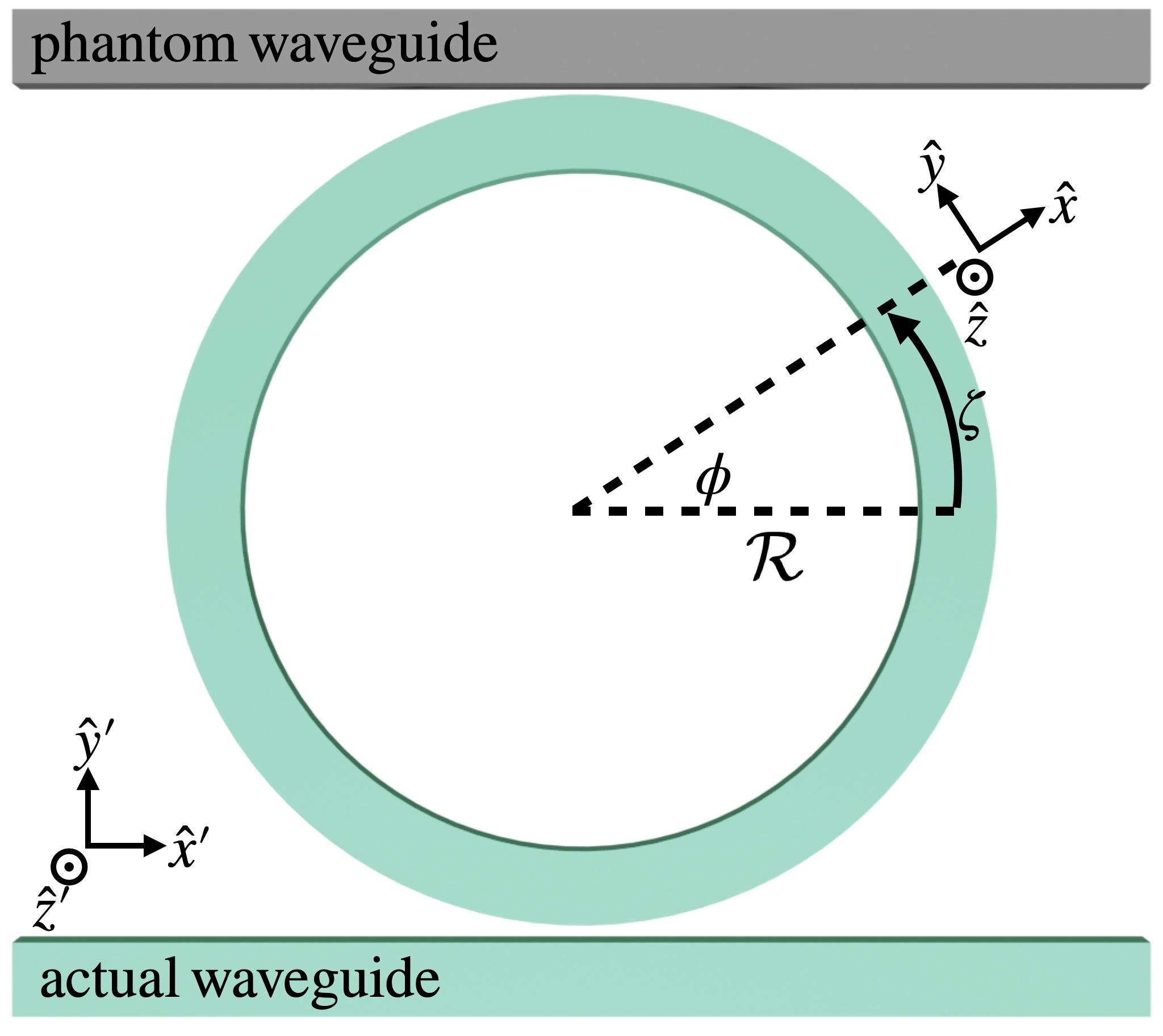}
    \caption{Microring coupled to an actual and phantom waveguide.}
    \label{fig:Struct}
\end{figure}
We envision a pump pulse or cw excitation coupled from the waveguide into the ring at frequencies within one of the ring resonances, and photon pairs either {generated at frequencies within different ring resonances} {(nondegenerate SPDC, considered in the text), or} within one resonance (degenerate SPDC, {considered in Appendix \ref{app:DegenerateCase}}).

\subsection{\label{Hamiltonians}{The linear Hamiltonian}}

The linear Hamiltonian of our system consists of the sum of the Hamiltonians of its components and the Hamiltonians describing the coupling between them. We begin with an isolated ring, for which the Hamiltonian is given by 
\begin{equation}
    H_{\mathrm{ring}} = \sum_u \hbar \omega_u b^\dagger_u b_u,
    \label{eq:RingHamil}
\end{equation} 
where the resonant frequencies are denoted by $\omega_u$, and the operators $b_u$ satisfy the commutation relations
\begin{equation}
[b_u,b_{u'}^\dagger] = \delta_{uu'}.
\label{eq:CommuatationRelations_bJ}
\end{equation}
The associated displacement field inside the ring is given by 
\begin{equation}
    \bm{\mathsf{D}}(\boldsymbol{r}) = \sum_u \sqrt{\frac{\hbar \omega_{u}}{2\mathcal{L}}} \bm{\mathsf{d}}_{u}(\boldsymbol{r}_\perp,\zeta) b_u e^{i\kappa_u \zeta} + \mathrm{H.c.}, 
    \label{eq:DFieldsInRing}
\end{equation}
where $\zeta$ indicates the distance along the circle at the nominal radius $\mathcal{R}$ of the ring, and $\boldsymbol{r}_\perp$ is a vector in the local $zx$ plane (see Fig. 1), $\kappa_u = 2\pi m_u / \mathcal{L}$ are the resonant wavenumbers in the ring with $m_u$ the associated mode numbers \cite{bib:Quesada2022}, of which the positive ones will be relevant here, and $\mathcal{L} = 2 \pi \mathcal{R}$; the $\bm{\mathsf{d}}_{u}(\boldsymbol{r}_\perp,\zeta)$ are the field amplitudes, normalized according to
\begin{equation}
\begin{split}
    \int \frac{v_p(\boldsymbol{r}_\perp;\omega_{u})}{v_g(\boldsymbol{r}_\perp;\omega_{u})} \frac{\bm{\mathsf{d}}_{u}^* (\boldsymbol{r}_\perp) \cdot \bm{\mathsf{d}}_{u} (\boldsymbol{r}_\perp)}{\epsilon_0 \varepsilon_1(\boldsymbol{r}_\perp;\omega_u)}  \mathrm{d} \boldsymbol{r}_\perp = 1,
\end{split}
    \label{eq:NormalizationConditionDfieldsNEW}
\end{equation}
where $\varepsilon_1(\boldsymbol{r}_\perp,\omega_u)$ is the relative permittivity, $v_g(\boldsymbol{r}_\perp;\omega_{u})$ is the local group velocity of the material,
and $v_p(\boldsymbol{r}_\perp;\omega_{u})$ is the local phase velocity of the material, all at frequency $\omega_u$ \cite{bib:Quesada2022}. Here we have used the fact that the dot product $\bm{\mathsf{d}}_{u}^* (\boldsymbol{r}_\perp,\zeta) \cdot \bm{\mathsf{d}}_{u} (\boldsymbol{r}_\perp,\zeta)$ is independent of $\zeta$, and we have written it simply as $\bm{\mathsf{d}}_{u}^* (\boldsymbol{r}_\perp) \cdot \bm{\mathsf{d}}_{u} (\boldsymbol{r}_\perp)$. {In the systems treated here we consider pump frequencies near a ring resonance frequency that we denote by $\omega_{P}$, signal frequencies near a ring resonance that we denote by $\omega_{S}$, and idler frequencies near a ring resonance that we denote by $\omega_{I}$; for each pump scenario we also consider {a range of}  different $\omega_{S}$ and $\omega_{I}$.} 

Next we consider an isolated waveguide, and focus first on the actual waveguide in Fig. 1 in this isolated limit. We write the total displacement field as
\begin{equation}
    \boldsymbol{D}(\boldsymbol{r}) = \sum_u \boldsymbol{D}_u(\boldsymbol{r}) + \mathrm{H.c.},
    \label{eq:TotalDisplacement}
\end{equation}
where here each $u$ labels a frequency ``bin" with centre at the ring resonance $\omega_u$, and extending over all frequencies of light in the waveguide relevant for coupling into that ring resonance. So we have 
\begin{equation}
    \boldsymbol{D}_{u}(\boldsymbol{r}) =\int \mathrm{d}k \boldsymbol{D}_{uk}(\boldsymbol{r}) a_u(k),
    \label{eq:DispWithLadder}
\end{equation}
and the integral ranges over the $k$ in frequency bin $u$. The ladder operators satisfy the usual commutation relations \cite{bib:Quesada2022}
\begin{equation}
[a_u(k),a_{u}^\dagger(k')] = \delta(k - k'),
\label{eq:CommuatationRelations_aJ}
\end{equation}
{with} all the other commutators {vanishing,} {and}
\begin{equation}
    \boldsymbol{D}_{uk}{(\boldsymbol{r})} = \sqrt{\frac{\hbar \omega_{uk}}{4\pi}} \boldsymbol{d}_{uk}(\boldsymbol{r}_\perp) e^{iks},
    \label{eq:FieldsWG}
\end{equation}
where $\boldsymbol{r}_\perp$ is the vector in a cross-section of the waveguide, perpendicular to the direction of propagation, which is indicated by increasing $s$; for the actual waveguide in Fig. 1, these correspond respectively to the $y'z'$ plane and the direction $x'$. In analogy with Eq. \eqref{eq:NormalizationConditionDfieldsNEW}, the field amplitudes are normalized \cite{bib:Quesada2022} according to 
\begin{equation}
\begin{split}
    \int \frac{v_p(\boldsymbol{r}_\perp;\omega_{uk})}{v_g(\boldsymbol{r}_\perp;\omega_{uk})} \frac{{\boldsymbol{d}}_{uk}^* (\boldsymbol{r}_\perp)  \cdot  \boldsymbol{d}_{uk} (\boldsymbol{r}_\perp)}{\epsilon_0 \varepsilon_1(\boldsymbol{r}_\perp,\omega_{uk})}  \mathrm{d} \boldsymbol{r}_\perp = 1.\\
\end{split}
    \label{eq:NormalizationConditionDfields}
\end{equation}
Here we have written the frequency associated with wavenumber $k$ within bin $u$ as 
\begin{equation}
     \omega_{uk} = \omega_u + v_u(k-K_u),
    \label{eq:OmegaFuncK}
\end{equation} 
where group velocity dispersion and higher-order terms are neglected for the small frequency ranges in each frequency bin $u$; the group velocity associated with bin $u$ is denoted by $v_u=\left(\partial\omega_{uk}/\partial k\right)_{K_u}$, with $K_u$ the value of $k$ at the centre frequency $\omega_u$. 
It is then convenient to introduce a channel operator associated with each frequency bin,
\begin{equation}
    \psi_u(s) = \int \frac{\mathrm{d}k}{\sqrt{2\pi}} a_u(k)e^{i(k-K_u)s},
    \label{eq:psi(s)}
\end{equation}
and using Eq. \eqref{eq:OmegaFuncK} the Hamiltonian of the waveguide can be written \cite{bib:Quesada2022} as 
\begin{equation}
\begin{split}
    H_{\mathrm{wg}} &= \sum_u \Bigg[\int \hbar \omega_u \psi^\dagger_u(s) \psi_u(s) \mathrm{d}s \\
    & - \frac{i\hbar v_u}{2} \int \left( \psi^\dagger_u(s) \frac{\partial \psi_u(s)}{\partial s} - \frac{\partial \psi^\dagger_u(s)}{\partial s}\psi_u(s) \right) \mathrm{d}s\Bigg]. 
    \label{eq:WGHamiltonian}
\end{split}
\end{equation}
We adopt a point coupling model between the waveguides and the ring resonator. For the actual waveguide, we take the coupling point at $s=0$ and the coupling Hamiltonian is given by
\begin{equation}
    H_{\mathrm{cpl}} = \sum_u (\hbar {\gamma}_u b^\dagger_u \psi_u(0) +\mathrm{H.c.}),
    \label{eq:CouplingHamiltonian}
\end{equation}
where $\gamma_u$ characterizes the strength of the coupling between a discrete ring mode and the associated waveguide field operator $\psi_u(s)$. This expression is valid in the high finesse regime \cite{bib:Banic2022}. 

Up to this point, only expressions for the actual waveguide have been introduced. {We can generalize these to refer to either the actual or the phantom waveguide by introducing an index} $\lambda\in\{\mathrm{ac},\mathrm{ph}\}$. Then for the actual (phantom) waveguide, the direction of increasing propagation is $s=x'$ ($s=-x'$), with the field operator {denoted by} $\psi_u^{\mathrm{ac}}(s)$ ($\psi_u^{\mathrm{ph}}(s)$) with the coupling constant at $s=0$ being $\gamma_u^{\mathrm{ac}}$ ($\gamma_u^{\mathrm{ph}}$). We rewrite Eqs. \eqref{eq:OmegaFuncK} and \eqref{eq:psi(s)} {as} 
\begin{equation}
    \omega_{uk}^\lambda = \omega_u + v_u^\lambda(k-K_u^\lambda),
    \label{eq:OmegaFuncKLam}
\end{equation}
{to apply to both waveguides,} where $v_u^\lambda=\left(\partial\omega_{uk}^\lambda/\partial k\right)_{K_u^\lambda}$, and
\begin{equation}
    \psi_u^\lambda(s) = \int \frac{\mathrm{d}k}{\sqrt{2\pi}} a_u^\lambda(k)e^{i(k-K_u^\lambda)s},
    \label{eq:psi(s)_Lam}
\end{equation}
and where the operators commute for different waveguides
\begin{equation}
\left[  a_{u}^{\lambda}(k), \left\{ a_{u}^{\lambda'}(k') \right\}^\dagger \right] = \delta_{\lambda\lambda'}\delta(k-k').
    \label{eq:CommutationRelationInOut}
\end{equation}
Using these definitions, the total linear Hamiltonian of our system in Fig. \ref{fig:Struct} is then
\begin{equation}
\begin{split}
    H_L &= H_{\mathrm{ring}} + H^{\mathrm{ac}}_{\mathrm{wg}} + H^{\mathrm{ph}}_{\mathrm{wg}}  + H^{\mathrm{ac}}_{\mathrm{cpl}} + H^{\mathrm{ph}}_{\mathrm{cpl}}.
    \label{eq:TotalHamiltonian}
    \end{split}
\end{equation}

\subsection{\label{asy-in-out}{Asymptotic fields}}
In calculating the nonlinear response of structures such as the ring resonator we consider here, one could of course proceed by using an expansion of the full displacement field in terms of the ``ring mode fields" from Eq. \eqref{eq:DFieldsInRing} and the ``waveguide mode fields" from Eq. \eqref{eq:TotalDisplacement}. But since there is linear coupling between these elements of the structure, that would complicate the analysis of nonlinear effects. An alternate strategy is to employ ``asymptotic-in mode fields" and ``asymptotic-out mode fields" \cite{bib:Liscidini2012}, an extension of the use of asymptotic-in and asymptotic-out states in scattering theory \cite{bib:Breit1954}.  
 \begin{figure}[ht]
    \centering
\includegraphics[width=0.48\textwidth]{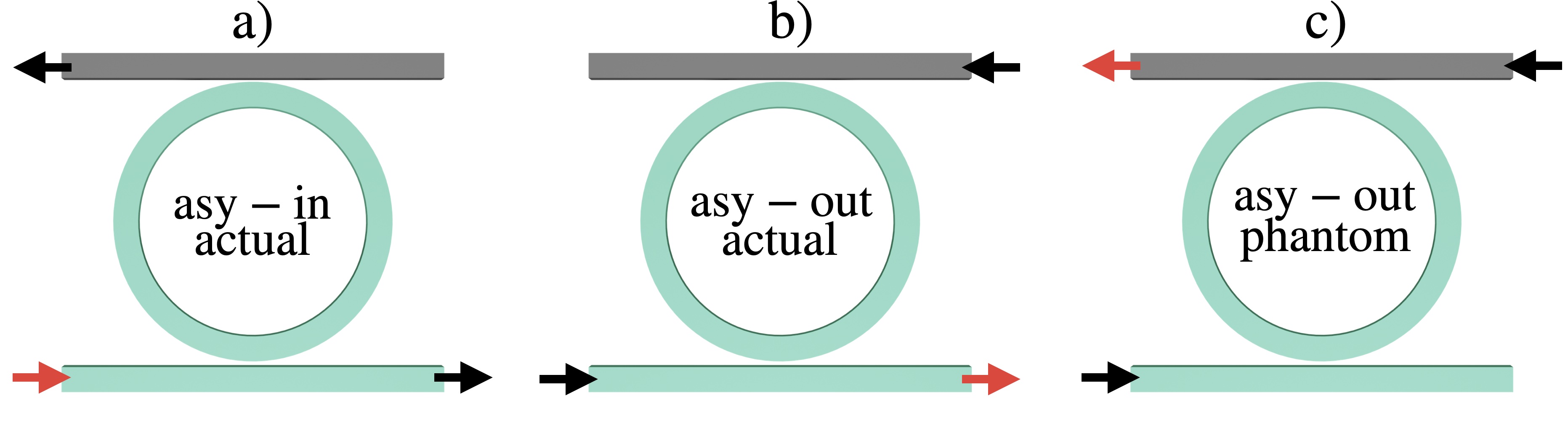}
    \caption{Schematic of the a) asymptotic-in {field} for the actual waveguide, b) asymptotic-out {field} for the actual waveguide, c) asymptotic-out {field} for the phantom waveguide.} 
    \label{fig:TripleRing}
\end{figure}

For the system we are considering here the asymptotic-in {mode fields form a complete set, as do the asymptotic-out mode fields,} so we can write either 
\begin{equation}
    \boldsymbol{D}(\boldsymbol{r}) = \sum_u \boldsymbol{D}_{u}^{\mathrm{out}}(\boldsymbol{r}) + \rm{H.c.}
    \label{eq:TotalAsyFields1}
\end{equation}
or
\begin{equation}
    \boldsymbol{D}(\boldsymbol{r}) = \sum_u \boldsymbol{D}_{u}^{\mathrm{in}}(\boldsymbol{r}) + \rm{H.c.}, 
    \label{eq:TotalAsyFields2}
\end{equation}
with 
\begin{equation}
\begin{split}
    \boldsymbol{D}_{u}^{\rm{out}}(\boldsymbol{r})&= \sum_\lambda \int \mathrm{d}k \boldsymbol{D}_{uk}^{\mathrm{out},\lambda} (\boldsymbol{r})  a_{u}^{\mathrm{out},\lambda}(k),\\
    \boldsymbol{D}_{u}^{\rm{in}}(\boldsymbol{r})&= \sum_\lambda \int \mathrm{d}k \boldsymbol{D}_{uk}^{\mathrm{in},\lambda} (\boldsymbol{r})  a_{u}^{\mathrm{in},\lambda}(k),
    \end{split}
    \label{eq:TotalAsyFields22}
\end{equation}
where the sum over $\lambda$ is summing over the different waveguides. Both the fields $\boldsymbol{D}_{uk}^{\mathrm{out},\lambda} (\boldsymbol{r})$ and $\boldsymbol{D}_{uk}^{\mathrm{in},\lambda} (\boldsymbol{r})$ are generally nonzero in the ring and in both waveguides, and in the absence of nonlinearity the Heisenberg operator versions of $a_{u}^{\mathrm{out},\lambda}(k)$ and $a_{u}^{\mathrm{in},\lambda}(k)$ are respectively $a_{u}^{\mathrm{out},\lambda}(k)e^{-i\omega_{uk}^\lambda t}$ and $a_{u}^{\mathrm{in},\lambda}(k)e^{-i\omega_{uk}^\lambda t}$. That is, the asymptotic-in and -out mode fields identify modes of the full linear Hamiltonian (Eq. \eqref{eq:TotalHamiltonian}).

The mode fields $\boldsymbol{D}_{uk}^{\mathrm{out},\lambda} (\boldsymbol{r})$ are constructed {(see Fig. 2b and 2c)} so that they are equal to $\boldsymbol{D}_{uk} (\boldsymbol{r})$ for waveguide $\lambda$ and $s>0$ {(indicated by the red arrows)}, and vanish for the other waveguide for $s>0$; the mode fields $\boldsymbol{D}_{uk}^{\mathrm{in},\lambda} (\boldsymbol{r})$ {(see Figure 2a for the asymptotic-in mode field for the actual waveguide)} are constructed so that they are equal to $\boldsymbol{D}_{uk} (\boldsymbol{r})$ for waveguide $\lambda$ and $s<0$ {(indicated by the red arrow)} and vanish for the other waveguide for $s<0$.  So the asymptotic-out expansion is appropriate for the signal and idler fields exiting the structure, while the asymptotic-in expansion is appropriate for the pump field entering the structure \cite{bib:Liscidini2012,bib:Banic2022,bib:Breit1954}. 

We assume that the nonlinearity is effective only in the ring resonator, where the fields are enhanced.  Thus, in constructing the nonlinear interaction Hamiltonian we need the expressions for the fields $\boldsymbol{D}_{uk}^{\mathrm{out},\lambda} (\boldsymbol{r})$ and $\boldsymbol{D}_{uk}^{\mathrm{in},\lambda} (\boldsymbol{r})$ in the ring.  We denote these by $\bm{\mathsf{D}}_{uk}^{\mathrm{out},\lambda} (\boldsymbol{r})$ and $\bm{\mathsf{D}}_{uk}^{\mathrm{in},\lambda} (\boldsymbol{r})$, and they are given by \cite{bib:Banic2022}:
\begin{equation}
\begin{split}
    &\bm{\mathsf{D}}^{\mathrm{out}, \lambda}_{uk}(\boldsymbol{r}) = - \sqrt{\frac{\hbar \omega_u}{4 \pi}} \bm{\mathsf{d}}_u (\boldsymbol{r}_\perp , \zeta)  {F}_{u +}^{\lambda} (k) e^{i\kappa_u \zeta},\\
    &\bm{\mathsf{D}}^{\mathrm{in}, \lambda}_{uk}(\boldsymbol{r}) = - \sqrt{\frac{\hbar \omega_u}{4 \pi}} \bm{\mathsf{d}}_u (\boldsymbol{r}_\perp , \zeta)  {F}_{u -}^{\lambda} (k) e^{i\kappa_u \zeta},\\
    \end{split}
    \label{eq:FieldEnchance_NewNotation}
\end{equation}
where
\begin{equation}
     {F}_{u \pm}^{ \lambda} (k) = \frac{1}{\sqrt{\mathcal{L}}} \left( \frac{({\gamma}_u^{\lambda})^*}{v_u^{\lambda}(K_u^{\lambda} - k)\pm i\bar{\Gamma}_u} \right)
    \label{eq:FieldEnchance_Definition}
\end{equation}
{describes the resonant enhancement,} with $\bar{\Gamma}_u$ the total decay rate of the resonator, which is related to the decay rates into the actual and phantom waveguides by
\begin{equation}
    \bar{\Gamma}_u = \sum_\lambda \Gamma_u^\lambda = \Gamma_u^{\mathrm{ac}} + \Gamma_u^{\mathrm{ph}}.
    \label{eq:GammaBarUDef}
\end{equation}
{In the point coupling model we adopt these} decay rates {follow from the} group {velocities} and coupling constant{s in} the Hamiltonian \cite{bib:Banic2022}:
\begin{equation}
    \Gamma_u^\lambda = \frac{|{\gamma}_u^\lambda|^2}{2 v_u^\lambda}.
    \label{eq:GammaUDef}
\end{equation}
We introduce the loaded quality factor $Q_u^{\mathrm{(load)}}$, which is related to the total decay rate $\bar{\Gamma}_u$ {of the ring resonance $u$,}
\begin{equation}
\begin{split}
     \bar{\Gamma}_u &= \frac{\omega_u}{2Q_u^{\mathrm{(load)}}},
\end{split}
    \label{eq:GammaBarQLoaded}
\end{equation}
and the quality factors associated {with {the coupling between the ring and} the waveguide ${\lambda}$ are given by}
\begin{equation}
\begin{split}
     {\Gamma}_u^\lambda & = \frac{\omega_u}
     {2Q_u^\lambda}.  
\end{split}
    \label{eq:GammaBarQ}
\end{equation}
{Since the phantom waveguide models loss,} $Q_u^{\mathrm{ph}}$ is the ``intrinsic quality factor," and $Q_u^{\mathrm{ac}}$ is the ``extrinsic quality factor," {and}
\begin{equation}
    \frac{1}{Q_u^{\mathrm{(load)}}} =  \frac{1}{Q_u^{\mathrm{ac}}} + \frac{1}{Q_u^{\mathrm{ph}}.}
    \label{eq:QRealtions}
\end{equation}

\subsection{\label{NL}{The nonlinear Hamiltonian}}

We assume that the only {relevant nonlinear response coefficient is} $\chi^{(2)}$. We neglect the third-order processes such as self-phase modulation (SPM) and cross-phase modulation (XPM). However, for the sample calculations we present in Sec. \ref{sec:NumResults} we find that the shifts in the pump, signal, and idler resonances can be considered negligible -- approximately two orders of magnitude smaller than the resonance linewidth -- for continuous-wave pump powers below $100\,{\rm \mu W}$ \cite{Vernon2015}.  

With {the} assumption that the nonlinearity is important only in the ring{,} where the fields are enhanced, the nonlinear Hamiltonian takes the form \cite{bib:Quesada2022}
\begin{equation}
    H_{NL} = -\frac{1}{3\epsilon_0} \int \Gamma^{(2)}_{i'j'k'} (\boldsymbol{r}) {\mathsf{D}}^{i'}(\boldsymbol{r}) {\mathsf{D}}^{j'}(\boldsymbol{r}) {\mathsf{D}}^{k'}(\boldsymbol{r}) \mathrm{d} \boldsymbol{r},
    \label{eq:NonLinHamil_Chi2}
\end{equation}
where ${\mathsf{D}}^{l'}(\boldsymbol{r})$ is the $l'^{\mathrm{th}}$ Cartesian field component inside the ring, {the integration ranges over the ring, and for} frequencies around $\omega_P$, $\omega_S$, and $\omega_I$, we {have} 
\begin{equation}
    \Gamma^{(2)}_{i'j'k'}(\boldsymbol{r}) = \frac{\chi^{(2)}_{i'j'k'}(\boldsymbol{r})}{\epsilon_0 \varepsilon_1 (\boldsymbol{r};\omega_S) \varepsilon_1 (\boldsymbol{r};\omega_I) \varepsilon_1 (\boldsymbol{r};\omega_P)},
    \label{eq:Gamma2Chi2}
\end{equation}
{where $\chi^{(2)}_{i'j'k'}(\boldsymbol{r})$ is the nonlinear susceptibility at the frequencies of interest} \cite{bib:Quesada2022}.

We note here that the primed indices $\{i',j',k'\}$ {refer to} the laboratoryframe, and the unprimed indices $\{i,j,k\}$ to the ring frame, as described in Appendix \ref{app:RefFrames} and {indicated} in Fig. \ref{fig:Struct}. {The relevant term for the generation of} a pair of nondegenerate signal and idler photons, inside frequency bins $S$ and $I$ (where we {take} $\omega_S>\omega_I$), {from} a pump photon inside the bin labelled $P$ is
\begin{equation}
\begin{split}
    H^{\mathrm{SPDC}}_{NL} ={-} \frac{2}{\epsilon_0} \int \Gamma^{(2)}_{i'j'k'}(\boldsymbol{r}) & \left[{\mathsf{D}}_{S}^{\mathrm{out}(i')}(\boldsymbol{r}) {\mathsf{D}}_{I}^{\mathrm{out}(j')}(\boldsymbol{r})\right]^{\dagger} \\
    & \times {\mathsf{D}}_{P}^{\mathrm{in}(k')}(\boldsymbol{r}) \mathrm{d} \boldsymbol{r}
    + \mathrm{H.c.},
\end{split}
    \label{eq:H_NL_SPDC_NewSum}
\end{equation} 
where the displacement operators $\bm{\mathsf{D}}_u^{\mathrm{in/out}}(\boldsymbol{r})$ are those from Eq. \eqref{eq:TotalAsyFields22}, and we have used the expression for the displacement fields $\bm{\mathsf{D}}^{\mathrm{in/out}, \lambda}_{uk}(\boldsymbol{r})$ inside the ring from Eq. \eqref{eq:FieldEnchance_NewNotation}. The pump photons enter the ring from the actual waveguide, and the generated signal and idler photons exit the ring either by the phantom or the actual waveguide, depending on the label $\lambda$ and $\lambda'$. Using Eq. \eqref{eq:FieldEnchance_NewNotation} in Eq. \eqref{eq:H_NL_SPDC_NewSum} we obtain
\begin{equation}
\begin{split}
    H^{\mathrm{SPDC}}_{NL} & = - \sum_{\lambda\lambda'} \int  \mathrm{d} k_1 \mathrm{d} k_2 \mathrm{d} k_3 \mathsf{K}_{\lambda\lambda'}({k}_1,{k}_2,{k}_3) \\
    & \times \left[a_{S}^{\mathrm{out},\lambda}(k_1) a_{I}^{\mathrm{out},\lambda'}(k_2)\right]^\dagger a_{P}^{\mathrm{in},\mathrm{ac}}(k_3) + \mathrm{H.c.},
    \label{eq:H_NL}
    \end{split}
\end{equation}
where
\begin{equation}
\begin{split}
    \mathsf{K}_{\lambda\lambda'}(k_1,k_2,k_3) = & \sqrt{\frac{\hbar \omega_P}{4 \pi} \frac{\hbar \omega_{S}}{4 \pi} \frac{\hbar \omega_{I}}{4 \pi}}   \left[ F_{S+}^{\lambda}(k_1) F_{I+}^{\lambda'}(k_2) \right]^* \\
    & \times F_{P-}^{\mathrm{ac}}(k_3) \times \bar{\mathsf{K}}_{SIP},
    \label{eq:K_Expression}
    \end{split}
\end{equation}
and
\begin{equation}
\begin{split}
    \bar{\mathsf{K}}_{SIP}  =  2\epsilon_0 \int \chi^{(2)}_{i'j'k'}(\boldsymbol{r}) & [ {\mathsf{e}}_{S}^{i'}(\boldsymbol{r}) {\mathsf{e}}_{I}^{j'}(\boldsymbol{r}) ]^* {\mathsf{e}}_{P}^{k'}(\boldsymbol{r}) \\
    & \times e^{i(\kappa_P - \kappa_{S} - \kappa_{I})\zeta} \mathrm{d} \boldsymbol{r}_\perp  \mathrm{d} \zeta.
\end{split}
    \label{eq:K_bar_RingFrame}
\end{equation}
To derive Eq. \eqref{eq:K_bar_RingFrame} we have used the relation between the displacement and electric fields \cite{bib:Quesada2022},
\begin{equation} 
    \bm{\mathsf{d}}_u(\boldsymbol{r}) = \epsilon_0 \varepsilon_1(\boldsymbol{r};\omega_u) \bm{\mathsf{e}}_u(\boldsymbol{r}_\perp,\zeta).
    \label{eq:Electric_Displacement_Small}
\end{equation}
{In the ring frame, at points inside the ring} the elements of $\chi^{(2)}_{ijk}(\boldsymbol{r})$ for a zincblende material are given {by}
\begin{equation}
    \begin{split}
        \chi^{(2)}_{[xxz]}(\boldsymbol{r})&=\bar{\chi}^{(2)}\sin{(2\phi)},\\\chi^{(2)}_{[yyz]}(\boldsymbol{r})&=-\bar{\chi}^{(2)}\sin{(2\phi)},\\
        \chi^{(2)}_{[xyz]}(\boldsymbol{r})&=\bar{\chi}^{(2)}\cos{(2\phi)},
    \end{split}
    \label{eq:Chi2SinCos_InText}
\end{equation}
{(see Fig. 1), where $\bar{\chi}^{(2)}$ is the single parameter that characterizes the second-order response (see Appendix \ref{app:RefFrames}), and the} notation $[abc]$ denotes all distinct permutations of $\{a,b,c\}$. {As usual, we have assumed the crystal $z$ axis is perpendicular to the chip. Performing the} sums {in the expression from Eq. \eqref{eq:K_bar_RingFrame} for $\bar{\mathsf{K}}_{SIP}$, we find} \begin{equation}
    \begin{split}
        \bar{\mathsf{K}}_{SIP} =\epsilon_0 \bar{\chi}^{(2)}\Big[ V_{SIP}^{(+)} & \int_{\rm{ring}} \mathrm{d}  \boldsymbol{r}_\perp W_{SIP}^{(+)}(\boldsymbol{r}_\perp) \\
         &+ V_{SIP}^{(-)} \int_{\rm{ring}} \mathrm{d} \boldsymbol{r}_\perp W_{SIP}^{(-)}(\boldsymbol{r}_\perp)\Big],
    \end{split}
    \label{eq:KbarSI_WS}
\end{equation}
where
\begin{equation}
    \begin{split}
        W_{SIP}^{(\pm)}(\boldsymbol{r}_\perp) &= \Bigg\{ \sum\limits_{ i,j,k = [xyz]} [ {\mathsf{e}}_{S}^i(\boldsymbol{r}_\perp) {\mathsf{e}}_{I}^j(\boldsymbol{r}_\perp) ]^* {\mathsf{e}}_{P}^k(\boldsymbol{r}_\perp) \\
        & \mp \Bigg[ \sum\limits_{i,j,k = [xxz]} [ {\mathsf{e}}_{S}^i(\boldsymbol{r}_\perp) {\mathsf{e}}_{I}^j(\boldsymbol{r}_\perp) ]^* {\mathsf{e}}_{P}^k(\boldsymbol{r}_\perp) \\
        & + \,\,\,\,\,\sum\limits_{ i,j,k = [yyz]} [ {\mathsf{e}}_{S}^i(\boldsymbol{r}_\perp) {\mathsf{e}}_{I}^j(\boldsymbol{r}_\perp) ]^* {\mathsf{e}}_{P}^k(\boldsymbol{r}_\perp) \Bigg]\Bigg\},
    \end{split}
    \label{eq:KbarSI_WS_Developped}
\end{equation}
{with the notation ${i,j,k = [abc]}$ under the sum denoting that the $(i,j,k)$ triplet is summed only over the distinct permutations of $\{a,b,c\}$, and}

\begin{equation}
\begin{split}
    V_{SIP}^{(\pm)} &= \int_0^{\mathcal{L}} e^{i(\kappa_P - \kappa_{S} - \kappa_{I} \pm \frac{2}{\mathcal{R}})\zeta} \mathrm{d} \zeta \\
    &= 2\mathcal{R} e^{i \pi \mathcal{R} \Delta \kappa} \frac{\sin(\pi (\mathcal{R} \Delta \kappa \pm 2))}{\mathcal{R} \Delta \kappa \pm 2},
    \label{eq:f_sigma}
    \end{split}
\end{equation}
{with} $\Delta \kappa = \kappa_P - \kappa_{S} - \kappa_{I}$. {Since 
\begin{equation}
\mathcal{R} \Delta \kappa \pm 2 = m_P - m_{S} - m_{I} \pm 2,
    \label{eq:QuasiPhaseMatch_InText}
\end{equation}
which is an integer, we see that 
$V_{SIP}^{(\pm)}$ vanishes unless
\begin{equation}
m_P - m_{S} - m_{I} \pm 2=0,
\label{eq:QPMuse}
\end{equation}
which are the possible quasi-phase-matching conditions.} {Further, at most one of $V_{SIP}^{(\pm)}$ can vanish;} when the $+$ ($-$) quasi-phase-matched condition is met, the value of $V_{SIP}^{(+)}$ ($V_{SIP}^{(-)}$) is equal to $2\pi\mathcal{R}$ ($2\pi\mathcal{R}$) and the $V_{SIP}^{(-)}$ ($V_{SIP}^{(+)}$) term is zero, and so at most one term in Eq. \eqref{eq:KbarSI_WS} contributes to $\bar{\mathsf{K}}_{SIP}$. 
We write the “$\pm$ matched” version of Eq. \eqref{eq:K_bar_RingFrame} by combining Eq. \eqref{eq:KbarSI_WS} and the quasi-phase-matched $V_{SIP}^{(\pm)}$ terms,
\begin{equation}
\begin{split}
     \bar{\mathsf{K}}_{SIP}^{(\pm\,\mathrm{matched})} &= 2\epsilon_0 \bar{\chi}^{(2)} \pi \mathcal{R} \int_{\rm{ring}} \mathrm{d} \boldsymbol{r}_\perp W_{SIP}^{(\pm)}(\boldsymbol{r}_\perp).
\end{split}
    \label{eq:K_bar_RingFrameMatched}
\end{equation}  

To capture the effective area of the waveguide mode that is relevant for the nonlinear interaction, we can define 
\begin{equation}
    \begin{split}
        {A_\mathrm{eff}^{(\pm)}} &= \frac{{{N_{S} N_{I} N_P}}}{\left|\int_{\mathrm{ring}} \mathrm{d} \boldsymbol{r}_\perp W_{SIP}^{(\pm)}(\boldsymbol{r}_\perp) \right|^2},
    \end{split}
    \label{eq:AeffDefintion}
\end{equation}
where 
\begin{equation}
    N_u  = \int \frac{n(\boldsymbol{r}_\perp;\omega_u)/\bar{n}_u}{v_g(\boldsymbol{r}_\perp;\omega_u)/\bar{v}_u} \bm{\mathsf{e}}_u^* (\boldsymbol{r}_\perp) \cdot\bm{\mathsf{e}}_u (\boldsymbol{r}_\perp) \mathrm{d} \boldsymbol{r}_\perp,
    \label{eq:N_u}
\end{equation}
{with $\bar{v}_u$ and $\bar{n}_u$ typical values of the group velocities and group
refractive indices, respectively, introduced here just for convenience. Note that Eqs. \eqref{eq:AeffDefintion} and \eqref{eq:N_u} can be used regardless of the normalization of the waveguide fields, but if they are normalized according to Eq. \eqref{eq:NormalizationConditionDfieldsNEW} we immediately have $N_u=\bar{v}_u/(c \epsilon_0 \bar{n}_u)$. In terms of the effective area from Eq. \eqref{eq:AeffDefintion}, we can then rewrite Eq. \eqref{eq:K_bar_RingFrameMatched} as }     

\begin{equation}
\begin{split}
    \bar{\mathsf{K}}_{SIP}^{(\pm\,\mathrm{matched})} &= \frac{2\bar{\chi}^{(2)} \pi \mathcal{R}}{\epsilon_0^{1/2} c^{3/2}}  \sqrt{\frac{\bar{v}_{S} \bar{v}_{I} \bar{v}_P}{\bar{n}_{S} \bar{n}_{I} \bar{n}_P}} \frac{1}{\sqrt{A_{\mathrm{eff}}^{(\pm)}}}. \\
\end{split}
    \label{eq:K_bar_Leff_Relation}
\end{equation}

\subsection{\label{subsec:cw}{Continuous-wave excitation}}

{As a first calculation -- and to {help} identify the target parameters for the system -- we consider a monochromatic classical pump at frequency $\omega_0$, sufficiently weak that  lowest order perturbation theory can be applied and the pump can be considered undepleted. Then} we {can replace the pump annihilation operator by an amplitude} \cite{bib:Banic2022}, 
\begin{equation}
    a_P(k_3) \xrightarrow{} \alpha_P(k_3) = \sqrt{\frac{2 \pi P_P}{\hbar \omega_0 v_P^{\mathrm{ac}}}} \delta(k_3 - k_0),
    \label{eq:ClassicalPumP}
\end{equation}
where $P_P$ is the pump power, $v_P^{\mathrm{ac}}$ is the pump group velocity in the actual waveguide, and $\omega_0$ and $k_0$ are the frequency and wavevector associated with the cw pump respectively; the pump frequency $\omega_0$ can be detuned from the pump resonance frequency of the ring $\omega_P$. {With this substitution and moving into the interaction picture, from} Eq. \eqref{eq:H_NL} {we find}

\begin{equation}
\begin{split}
    H^{\mathrm{SPDC}}_{NL} (t)   &\rightarrow   \mathcal{H}^{\mathrm{}}_{NL} (t)  = -  \sum_{\lambda\lambda'}  \int \mathrm{d} k_1 \mathrm{d}  k_2 \mathsf{M}_{\lambda\lambda'} (k_1,k_2) \\
    & \times\left[a_{S}^{\mathrm{out},\lambda}(k_1) a_{I}^{\mathrm{out},\lambda'}(k_2)\right]^\dagger e^{-i\Omega_{SI}(k_1,k_2)t} + \mathrm{H.c.},\\
\end{split}
    \label{eq:H_NL_Classical}
\end{equation}
where 
\begin{equation}
\begin{split}
    \Omega_{SI}(k_1,k_2) \equiv \omega_0 - \omega_{Sk_1}^{\lambda} - \omega_{Ik_2}^{\lambda'} = A - v_{S}^{\lambda} k_1 - v_{I}^{\lambda'} k_2,
    \end{split}
    \label{eq:BigOmegaDefinition}
\end{equation}
with 
\begin{equation}
\Delta \omega \equiv \omega_{0} - \omega_{S} - \omega_{I}, \label{eq:DeltaOmega}
\end{equation}
$A \equiv \Delta \omega + v_{S}^\lambda K_{S}^{\lambda} + v_{I}^{\lambda'} K_{I}^{\lambda'}$, {and} 
\begin{equation}
    \mathsf{M}_{\lambda\lambda'} (k_1,k_2) \equiv \sqrt{\frac{2 \pi P_P}{\hbar \omega_0 v_P^{\mathrm{ac}}}}  \mathsf{K}_{\lambda\lambda'} (k_1,k_2,k_0). 
    \label{eq:MDefinition}
\end{equation}
{\subsubsection{Rates}}
As shown {earlier} \cite{bib:Banic2022}, {using} Eq. \eqref{eq:H_NL_Classical}, we {can} calculate the generated rate of photon pairs using Fermi's golden rule for a pump that is sufficiently weak. {The resulting expression for the} rate of generated photon pairs in distinct frequency bins labelled $S$ and $I$ {is}
\begin{equation}
    R^{SI} = \sum_{\lambda\lambda'} R^{SI}_{\lambda\lambda'}
    \label{eq:Rate_Pairs}
\end{equation}
where
\begin{equation}
R^{SI}_{\lambda\lambda'} = \frac{2 \pi}{\hbar^2} \int \mathrm{d} k_1 \mathrm{d} k_2 \delta(\Omega_{SI}(k_1,k_2)) |\mathsf{M}_{\lambda\lambda'} (k_1,k_2)|^2   
\label{eq:basic_rate}
\end{equation}
is the rate of generation of photon pairs where the signal photon ($S$) is exiting through the $\lambda$ waveguide and the idler photon ($I$) is exiting through the $\lambda'$ waveguide.
After integrating over both $k_1$ and $k_2$, we find
\begin{equation}
\begin{split}
    R^{SI}_{\lambda\lambda'} = & \frac{ P_P P_{\rm{vac}} \left(\bar{\chi}^{(2)}\right)^2 \sqrt{\omega_S\omega_I} \bar{\Gamma}_P \eta_{S}^{\lambda} \eta_{I}^{\lambda'} \eta_P^{\rm{(ac)}} }{2 \hbar \pi \epsilon_0 c^3 \bar{\Gamma}_S \bar{\Gamma}_I \mathcal{R}} \\
    & \,\,\,\,\,\,\,\,\,\,\,\,\,\,\,\,\,\,\,\, \times {\frac{\bar{v}_{S} \bar{v}_{I} \bar{v}_P}{\bar{n}_{S} \bar{n}_{I} \bar{n}_P}} \frac{1}{A_{\mathrm{eff}}^{(\pm)}} \frac{1}{\left[(\delta\omega)^2+(\bar{\Gamma}_P)^2\right]},
\end{split} \label{eq:Rate_Pairs_Total_ResonanceQuality_QuasiPhaseMatched}
\end{equation}
{which scales linearly with the pump power $P_P$,} where we have defined the {pump} detuning $\delta\omega\equiv\omega_0 - \omega_P$ and the escape efficiencies $\eta_u^\lambda$,
\begin{equation} \eta_u^\lambda\equiv\Gamma_u^\lambda / \bar{\Gamma}_u.
\label{eq:escape efficiencies}
\end{equation}
We note that the rate {expression from Eq. \eqref{eq:Rate_Pairs_Total_ResonanceQuality_QuasiPhaseMatched}} is for a given set of resonances {with resonant wavenumbers satisfying} the quasi-phase-matching condition given by Eq. \eqref{eq:QuasiPhaseMatch_InText}; if this condition is not met, {within our treatment of the coupling of the resonator to the channel} the generation rate {strictly vanishes.} The value of the effective area that should be used in the rate equation (either $A_{\mathrm{eff}}^{(+)}$ or $A_{\mathrm{eff}}^{(-)}$) is chosen by which quasi-phase-matching condition is met from Eq. \eqref{eq:QuasiPhaseMatch_InText} (either $+$ or $-$). We {have written} the rate as a function of the vacuum power \cite{Helt:12,bib:Banic2022}:
\begin{equation}
    P_{\rm{vac}} = \frac{\hbar}{2}\frac{\sqrt{\omega_S\omega_I}\bar{\Gamma}_S \bar{\Gamma}_I (\bar{\Gamma}_S + \bar{\Gamma}_I)}{(\Delta \omega)^2 + (\bar{\Gamma}_S + \bar{\Gamma}_I)^2}.
    \label{eq:VacPower}
\end{equation}
The rates for a different combination of exit waveguides $\mu$ and $\mu'$, are easily deduced from Eq. \eqref{eq:Rate_Pairs_Total_ResonanceQuality_QuasiPhaseMatched} \cite{bib:Banic2022},
\begin{equation}
    \frac{R^{SI}_{\lambda\lambda'}}{R^{SI}_{\mu\mu'}}
    =\frac{\eta_{S}^{\lambda} \eta_{I}^{\lambda'}}{\eta_{S}^{\mu} \eta_{I}^{\mu'}} = \frac{\Gamma_S^{\lambda} \Gamma_I^{\lambda'}}{\Gamma_S^{\mu} \Gamma_I^{\mu'}}.
    \label{eq:RatesDifferentExits}
\end{equation}
Of central interest will be the total rate at which both signal and idler photons leave the actual channel, 
\begin{equation}
R^{\mathrm{total}}_{\mathrm{ac},\mathrm{ac}}=
\hat R^{SS}_{\mathrm{ac,ac}} + \sum_{S>I} R^{SI}_{\mathrm{ac,ac}}
\label{eq:Rworkout}
\end{equation}
where $S$ and $I$ label the distinct frequency bins, {with} $S>I$ {indicating} that the centre frequency of the signal bin ($\omega_S$) is {taken} greater than the centre frequency of the idler bin ($\omega_I$). The expression for the nondegenerate rate $R^{SI}_{\mathrm{ac,ac}}$ {follows from} Eq. \eqref{eq:basic_rate}, and the degenerate efficiency $\hat R^{SS}_{\mathrm{ac,ac}}$ is {that of} Eq. \eqref{eq:Rate_PairsDegen}. {We can also introduce a generation efficiency $\mathcal{E}^{\mathrm{total}}_{\mathrm{ac,ac}}$, defined as the ratio of the rate at which photon pairs are generated in the actual channel to the rate at which pump photons are incident, 
\begin{equation}   
\mathcal{E}^{\mathrm{total}}_{\mathrm{ac,ac}}={R}^{\mathrm{total}}_{\mathrm{ac},\mathrm{ac}} \frac{\hbar\omega_0}{P_P}.
\label{eq:efficiency}
\end{equation}} 

{\subsubsection{Discussion}}
We present numerical results in Sec. \ref{sec:NumResults}, but we briefly mention {here} how certain parameters affect the rate of pair generation {and thus the efficiency}. 

{First, the rate} is inversely proportional to the effective area, {and so} it scales with the square of the integral in Eq. \eqref{eq:K_bar_RingFrameMatched}. Maximizing this integral {involves identifying the best} overlap between {the} electric-field mode profile{s.} 

{Second, we note} that the rate is inversely proportional to the radius of the ring $\mathcal{R}$, implying that lowering the radius of the ring might improve rates, although of course we can expect the intrinsic $Q$ to decrease as lower radii are considered due to increased scattering loss \cite{Thiel24}. 

Third, consider an $R^{SI}_{\lambda\lambda'}$ that does not vanish; this arises when the quasi-phase-matching condition from Eq. \eqref{eq:QPMuse} is satisfied. If the pump frequency were set to the ring resonance frequency $\omega_P$, and we wanted the signal and idler generated at the ring resonance frequencies $\omega_S$ and $\omega_I$ respectively, then from energy conservation we would require $\omega_P-\omega_S-\omega_I=0$. In general this cannot be satisfied together with the quasi-phase-matching condition{, which is required to have $R^{SI}_{\lambda\lambda'}\neq0$}. So the signal and the idler will not be generated exactly at the ring resonance frequencies, which will weaken the generation rate.

{More generally, if the pump frequency were detuned from the ring resonance $\omega_P$, set to a value $\omega_0 \neq \omega_P$ but still within the resonance centred at $\omega_P$, the pump intensity in the ring would be lower than 
if the excitation were at $\omega_P$, and this in itself would decrease the generation rate, as indicated by the nonzero value of $\delta\omega$ in the expression from Eq. \eqref{eq:Rate_Pairs_Total_ResonanceQuality_QuasiPhaseMatched} for
$R^{SI}_{\lambda\lambda'}$. Then the condition for energy conservation, with the signal and idler generated at the ring resonance frequencies $\omega_S$ and $\omega_I$ respectively, would be $\Delta\omega \equiv \omega_0-\omega_S-\omega_I = 0$. This in general will not be satisfied, but the generation rate is largest when this is as small as possible, with the signal and idler light as close as possible to $\omega_S$ and $\omega_I$ respectively. This effect is captured by the $P_{\mathrm{vac}}$ term in $R^{SI}_{\lambda\lambda'}$ from Eq. {\eqref{eq:VacPower}}, which peaks as $\Delta\omega \rightarrow 0$.}

To maximize the rate, both $\Delta\omega$ and $\delta\omega$ should be as close to zero as possible; specifically, {to have a non-negligible rate}, we should have ${|\Delta\omega|}\lesssim\bar{\Gamma}_S + \bar{\Gamma}_I$ and ${|\delta\omega|}\lesssim\bar{\Gamma}_P$. The latter condition can be satisfied {by choosing the pump frequency.} The former condition, {which we call the \emph{frequency bin matching condition},} is a challenge to obtain, and {is more difficult as rings of higher finesse are considered.}  

\subsection{\label{subsec:Pulsed}Pulsed excitation}
{We now consider excitation with a pump pulse, {and formulate the problem in terms of ``input" and ``output" kets, $\ket{\psi_\mathrm{in}}$ and $\ket{\psi_\mathrm{out}}$ respectively \cite{Yang:2008}.  For a pulse incident on the structure, the input ket is the ket that would describe the state to which the incident pulse would evolve at $t=0$ were there no nonlinearity, and the output ket is the ket at $t=0$ that would evolve to the actual state at later times, including the effects of the nonlinearity, if there were no nonlinearity. We take the input ket to} be a coherent state,}
\begin{equation}
\begin{split}
    \ket{\psi_\mathrm{in}} = e^{(\alpha A^\dagger_P - \mathrm{H.c.})} \ket{\mathrm{vac}},
    \end{split}
    \label{eq:Asy-InKet2}
\end{equation}
where
\begin{equation}
\begin{split}
    A^\dagger_P = \int \mathrm{d}k_3 \phi(k_3)\left[a_{P}^{\mathrm{in},\mathrm{ac}}(k_3)\right]^\dagger, 
    \label{eq:SuperPumpOp2}
    \end{split}
\end{equation}
{is the supermode pump creation operator \cite{Yang:2008}.} {The} pulse distribution {function} $\phi(k_3)$ is normalized according to
\begin{equation}
    \int \left|\phi(k_3) \right|^2\mathrm{d}k_3 = 1,
    \label{eq:NormalizationInText}
\end{equation}
{and the expectation value of the} total number of photons in the pump {pulse} is $N_P\equiv|\alpha|^2$.

\subsubsection{\label{subsub:PairRegime}{pair regime}}
We first look at the {pair regime}, where the outgoing state approximately consists of only pairs of photons. In this case, the output ket can be written as \cite{Yang:2008,bib:Banic2022}
\begin{equation}
\begin{split}
    \ket{\psi_\mathrm{out}} & \approx \ket{\psi_\mathrm{in}} + \beta(A_{\mathrm{II}}^{\mathrm{out}})^\dagger\ket{\psi_\mathrm{in}},\\
    \end{split}
    \label{eq:OutKetFinalFirstOrder}
\end{equation}
where $|\beta|^2$ is the probability of generating a pair of photons. {The} pair generation operator {is}
\begin{equation}
\begin{split}
    (A_{\mathrm{II}}^{\mathrm{out}})^\dagger \equiv \sum_{\lambda\lambda'}  & \int \mathrm{d}k_1 \mathrm{d}k_2 \varphi^\mathrm{II}_{\lambda\lambda'}(k_1,k_2)\\
    &\times\left[{a}_{S}^{\mathrm{out},\lambda}(k_1) {a}_{I}^{\mathrm{out},\lambda'}(k_2)\right]^\dagger,
    \end{split}
    \label{eq:SuperA}
\end{equation}
which is written in terms of the {components of the biphoton wave function (BWF) $\varphi^{\mathrm{II}}_{\lambda\lambda'}(k_1,k_2)$} associated {with} the signal photon exiting the $\lambda$ waveguide and the idler photon exiting the $\lambda'$ waveguide, 
\begin{equation}
\begin{split}
    \varphi^{\mathrm{II}}_{\lambda\lambda'}(k_1,k_2) \equiv \frac{i \alpha}{\beta\hbar}& \int_{-\infty}^{+\infty} \mathrm{d}t \int \mathrm{d}k_3  \mathsf{K}_{\lambda\lambda'}(k_1,k_2,k_3;t) \phi(k_3),
\end{split}
    \label{eq:BiphotonWF}
\end{equation}
{where the first (i.e., $k_1$) and second (i.e., $k_2$) arguments in the function relate to the signal and idler respectively, {and range only over those resonances;} the function is} normalized according to
\begin{equation}
\sum_{\lambda\lambda'}\int\mathrm{d}k_1\mathrm{d}k_2 \left|\varphi^{\mathrm{II}}_{\lambda\lambda'}(k_1,k_2)\right|^2 = 1.
    \label{eq:NormBiphoton}
\end{equation}
Since the operators for the asymptotic-out fields commute with those for the asymptotic-in fields, Eq. \eqref{eq:OutKetFinalFirstOrder} for the output ket can be written as
\begin{equation}
\begin{split}
    \ket{\psi_\mathrm{out}} = e^{(\alpha A^\dagger_P - \mathrm{H.c.})} \ket{\psi},
    \end{split}
    \label{eq:OutKetFinalFirstOrderNEW}
\end{equation}
where $\ket{\psi}$ is the state of the generated photons, which we write in the pair regime as
\begin{equation}
\ket{\psi} \approx \ket{\mathrm{vac}} + \beta\sum_{\lambda\lambda'}\ket{\mathrm{II}_{\lambda\lambda'}}.
    \label{eq:PsiGen}
\end{equation}
The ket $\ket{\mathrm{II}_{\lambda\lambda'}}$ represents a two-photon state where the signal photon exits by waveguide $\lambda$ and the idler photon exits by waveguide $\lambda'$: 
\begin{equation}
\begin{split}
    \ket{\mathrm{II}_{\lambda\lambda'}} \equiv \int \mathrm{d}&k_1\mathrm{d}k_2\varphi^{\mathrm{II}}_{\lambda\lambda'}(k_1,k_2)\\
    & \times \left[a_{S}^{\mathrm{out},\lambda}(k_1) a_{I}^{\mathrm{out},\lambda'}(k_2)\right]^\dagger \ket{\mathrm{vac}}.
    \label{eq:BWFLambdaTwoKet}
    \end{split}
\end{equation}
The states {$\ket{\mathrm{II}_{\lambda\lambda'}}$} and $\ket{\mathrm{II}_{\mu\mu'}}$ are orthogonal  {for $\lambda \neq \mu$ and $\lambda' \neq \mu'$,}
\begin{equation}
\begin{split}
\braket{\mathrm{II}_{\lambda\lambda'}}{{\mathrm{II}_{\mu\mu'}}} 
=& \delta_{\lambda\mu} \delta_{\lambda'\mu'} \\
 & \times \int \mathrm{d} k_1\mathrm{d}k_2 \varphi^{\mathrm{II}}_{\lambda\lambda'}(k_1,k_2) \varphi^{\mathrm{II}*}_{\mu\mu'}(k_1,k_2).
    \label{eq:Deriv84_3}
    \end{split}
\end{equation}
We can write the full two-photon ket by summing over the output waveguides,
\begin{equation}
    \ket{\mathrm{II}} =  \sum_{\lambda\lambda'}\ket{\mathrm{II}_{\lambda\lambda'}},
    \label{eq:TwoPhotonState}
\end{equation}
{where} 
\begin{equation}
    \braket{\mathrm{II}}{\mathrm{II}} = 1.
    \label{eq:NormKet}
\end{equation}
Evaluating the two integrals in Eq. \eqref{eq:BiphotonWF} and neglecting group velocity dispersion over the pump bandwidth, we obtain an analytical expression for the {components of the} BWF
\begin{equation}
\begin{split}
    \varphi^{\mathrm{II}}_{\lambda\lambda'}(k_1,k_2)= \frac{2\pi i\alpha}{\beta {v_P^{\mathrm{ac}}} \hbar}  \mathsf{K}_{\lambda\lambda'}(k_1,k_2,X_P)   \phi \left(X_P\right),
    \end{split}
    \label{eq:BiphotonWFExpression}
\end{equation}
where
\begin{equation}
    X_P \equiv K_P^{\mathrm{ac}} - \frac{(\omega_P -\omega_{Sk_1}^{\lambda} -\omega_{Ik_2}^{\lambda'})}{v_P^{\mathrm{ac}}}.
    \label{eq:X_P}
\end{equation}
{In this {pair regime}, the probability $|\beta|^2$ that a pair of photons is generated is determined by using the expression from Eq. \eqref{eq:BiphotonWFExpression} for the components of the biphoton wave function in the normalization condition from Eq. \eqref{eq:NormBiphoton}. We find} 
\begin{equation}
\begin{split}
    |\beta|^2 & = \frac{4\pi^2 E_P}{(v_P^{\mathrm{ac}})^2\hbar^3\omega_0}\\
    & \,\,\,\,\,\,\,\,\,\, \times \sum_{\lambda\lambda'}\int\mathrm{d}k_1\mathrm{d}k_2 \left|  \mathsf{K}_{\lambda\lambda'}(k_1,k_2,X_P)   \phi \left(X_P\right)\right|^2,
    \end{split}
    \label{eq:PairsInPulse}
\end{equation}
where $E_P$ is the pump pulse energy; this
expression scales linearly with the energy of the pump pulse for a fixed pulse duration. We can define the probabilities from the different combinations of output waveguides
\begin{equation}
\begin{split}
    |\beta_{\lambda\lambda'}|^2 \equiv & \frac{4\pi^2 E_P}{(v_P^{\mathrm{ac}})^2\hbar^3\omega_0}\\
    & \times \int\mathrm{d}k_1\mathrm{d}k_2 \left|  \mathsf{K}_{\lambda\lambda'}(k_1,k_2,X_P)   \phi \left(X_P\right)\right|^2,
    \label{eq:BetaLambLamba}
    \end{split}
\end{equation}
where the total probability is given by
\begin{equation}
    |\beta|^2 = \sum_{\lambda\lambda'}|\beta_{\lambda\lambda'}|^2.
    \label{eq:sumBeta}
\end{equation}
Dividing by the number of pump photons, {the efficiency for the pair regime, which {does not vary} within this regime}, can be identified as
\begin{equation}
\begin{split}
    \mathcal{E}_\mathrm{pulse; pair regime}^{SI} & = \frac{|\beta|^2}{|\alpha|^2},
    \label{eq:LowGain}
    \end{split}
\end{equation}
and the efficiency for each output channel is
\begin{equation}
\begin{split}
    [\mathcal{E}_\mathrm{pulse; pair regime}^{SI}]_{\lambda\lambda'} & = \frac{|\beta_{\lambda\lambda'}|^2}{|\alpha|^2},
    \label{eq:LowGainEach}
    \end{split}
\end{equation}
and {similar} to Eq. \eqref{eq:RatesDifferentExits} we have 
\begin{equation}
\begin{split}
    \frac{|\beta_{\lambda\lambda'}|^2}{|\beta_{\mu\mu'}|^2} = \frac{[\mathcal{E}_\mathrm{pulse; pair regime}^{SI}]_{\lambda\lambda'}}{[\mathcal{E}_\mathrm{pulse; pair regime}^{SI}]_{\mu\mu'}} = \frac{\eta_{S}^{\lambda} \eta_{I}^{\lambda'}}{\eta_{S}^{\mu} \eta_{I}^{\mu'}} = \frac{\Gamma_S^{\lambda} \Gamma_I^{\lambda'}}{\Gamma_S^{\mu} \Gamma_I^{\mu'}}.
    \label{eq:Frac}
\end{split}
\end{equation}

\subsubsection{\label{subsub:HighGain}{Lowest-order squeezing regime}}
To explore the {lowest-order squeezing (LOS) regime}, where {we must consider amplitudes for the generation of more than one photon pair by the pump pulse,} we employ a ``backwards Heisenberg picture" approach \cite{Yang:2008}, {and perform} the calculation to first order in the operator dynamics {(hence ``lowest order squeezing"),} which {leads to} an undepleted pump approximation. The result is equivalent to {a more standard calculation} assuming a classical, undepleted pump from the start, and neglecting time-ordering corrections in the evolution operator. {Using either approach, we} find 
\begin{equation}
\begin{split}
    \ket{\psi_\mathrm{out}}=& e^{\left[\beta(A_{\mathrm{II}}^{\mathrm{out}})^\dagger -\mathrm{H.c.}\right]}\ket{\psi_\mathrm{in}},
    \end{split}
    \label{eq:OutKetFinal2}
\end{equation}
where {within the approximations identified above} the pair generation operator is the same as previously defined in Eq. \eqref{eq:SuperA}, {and the norm of $\beta$ is still given by Eq. \eqref{eq:PairsInPulse},} but now $\beta$ ({the} ``squeezing parameter" \cite{bib:Drago}), no longer {immediately gives} the probability of generating pairs, {and $\varphi^{\mathrm{II}}_{\lambda\lambda'}(k_1,k_2)$ is more properly identified as the ``joint spectral amplitude."} {The ket for the generated light in the LOS regime is given by a multimode squeezed vacuum state }
\begin{equation}
\ket{\psi} = e^{\left[\beta(A_{\mathrm{II}}^{\mathrm{out}})^\dagger -\mathrm{H.c.}\right]}\ket{\mathrm{vac}}. 
    \label{eq:GenKetSqueezed}
\end{equation}
{Using the state in Eq. \eqref{eq:GenKetSqueezed} we now calculate the number of squeezed photons} and other useful quantities such as the moments and the Schmidt number. To do this we discretize the wavevectors in the actual and phantom waveguides. We begin by discretizing 
Eq. \eqref{eq:SuperA}:
\begin{equation}
\begin{split}
    \left( A_\mathrm{II}^{\mathrm{out}}\right)^\dagger = \sum_{\lambda\lambda'} \sum_{ij} \sqrt{\Delta k_1 \Delta k_2}\Phi^{\mathrm{II}(ij)}_{\lambda\lambda'}\left[{a}_{Si}^{\mathrm{out},\lambda} {a}_{Ij}^{\mathrm{out},\lambda'}\right]^\dagger,
    \end{split}
    \label{eq:SuperAOutDiscrete}
\end{equation}
and we combine the indices $\lambda$ and $i$ to $m$, and the indices $\lambda'$ and $j$ to $m'$. The index discretizing $k_1$ ($k_2$) is $i$ ($j$) and $i=\left\{1,\dots,\ell_1\right\}$ ($j=\left\{1,\dots,\ell_2\right\}$), and since $\lambda$ ($\lambda'$) is summed over the two indices ``ac" and ``ph": $m$ ($m'$) is summed over $2\ell_1$ ($2\ell_2$) indices. We can then write Eq. \eqref{eq:SuperAOutDiscrete} as
\begin{equation}
\begin{split}
    \left( A_\mathrm{II}^{\mathrm{out}}\right)^\dagger = \sum_{mm'} \sqrt{\Delta k_1 \Delta k_2} \Phi^{\mathrm{II}}_{mm'}\left[{a}_{S}^{\mathrm{out},m} {a}_{I}^{\mathrm{out},m'}\right]^\dagger,
    \end{split}
    \label{eq:SuperAOutDiscrete2}
\end{equation}
and we define a $2\ell_1 \times 2\ell_2$ 
squeezing matrix $\boldsymbol{J}$ (\cite{Vendromin24}) with matrix elements
\begin{equation}
\begin{split}
    J_{mm'} = \beta \sqrt{\Delta k_1 \Delta k_2} \Phi^{\mathrm{II}}_{mm'}, 
    \end{split}
    \label{eq:JMatrix}
\end{equation}
which allows us to write Eq. \eqref{eq:SuperAOutDiscrete2} as
\begin{equation}
\begin{split}
    \left( A_\mathrm{II}^{\mathrm{out}}\right)^\dagger = \frac{1}{\beta} \sum_{mm'}  J_{mm'} \left[{a}_{S}^{\mathrm{out},m} {a}_{I}^{\mathrm{out},m'}\right]^\dagger,
    \end{split}
    \label{eq:SuperAOutDiscrete3}
\end{equation}
{and} to finally write our generated state as
\begin{equation}
\begin{split}
    \ket{\psi} =& \exp \left\{\sum_{mm'}  J_{mm'} \left[{a}_{S}^{\mathrm{out},m} {a}_{I}^{\mathrm{out},m'}\right]^\dagger -\mathrm{H.c.}\right\}\ket{\mathrm{vac}}.
    \end{split}
    \label{eq:OutKetFinal4}
\end{equation}
The matrix $\boldsymbol{J}$ is built from the four discretized {joint spectral amplitude} matrices:
\begin{equation}
    \boldsymbol{J} = \beta \sqrt{\Delta k_1 \Delta k_2} \left[\begin{matrix} \Phi^{\mathrm{II}}_{\mathrm{ac,ac}} & \Phi^{\mathrm{II}}_{\mathrm{ac,ph}} \\ \Phi^{\mathrm{II}}_{\mathrm{ph,ac}} & \Phi^{\mathrm{II}}_{\mathrm{ph,ph}}  
    \end{matrix} \right].
   \label{eq:BigMatrix}
\end{equation}
{The moments of the squeezed state {of} Eq. \eqref{eq:OutKetFinal4} are given by}
\begin{align}
\centering
    N_{mm'}^S &= \bra{\psi_\mathrm{out}}\left[{a}_{S}^{\mathrm{out},m}\right]^\dagger {a}_{S}^{\mathrm{out},m'} \ket{\psi_\mathrm{out}},\\
    N_{mm'}^I &= \bra{\psi_\mathrm{out}} \left[{a}_{I}^{\mathrm{out},m}\right]^\dagger {a}_{I}^{\mathrm{out},m'}\ket{\psi_\mathrm{out}},     \label{eq:NandMMoments}\\
    M_{mm'}^{SI} &= \bra{\psi_\mathrm{out}} {a}_{S}^{\mathrm{out},m} {a}_{I}^{\mathrm{out},m'} \ket{\psi_\mathrm{out}}. 
\end{align}
To calculate {these moments} we perform a Schmidt decomposition \cite{bib:Quesada2022, bib:Houde2023} of the {joint spectral amplitude} in Eq. \eqref{eq:OutKetFinal4}. This is achieved by decomposing the matrix $\boldsymbol{J}$ as
\begin{equation}
    \boldsymbol{J} = \boldsymbol{F}_S \boldsymbol{R} \left[\boldsymbol{F}_I\right]^\mathrm{T},
    \label{eq:SVD_J}
\end{equation}
where $\boldsymbol{F}_S$ and $\boldsymbol{F}_I$ are two square unitary matrices of rank $2\ell_1$ and $2\ell_2$ respectively{, and the} matrix $\boldsymbol{R}$ is a $2\ell_1\times2\ell_2$ diagonal matrix of rank $\ell = \min\left\{ 2\ell_1,2\ell_2 \right\}$ with entries $r_n$, for $n=\{1,\dots,\ell\}$. {Using Eq. \eqref{eq:SVD_J} for ${\bm J}$ it is straightforward to show \cite{bib:Quesada2022} that the moments in Eq. \eqref{eq:NandMMoments} can be written as}
\begin{align}
    N_{mm'}^S &= \left[\bm F_S \sinh^2(\bm R) \bm F^\dagger_S\right]_{mm'},
    \label{eq:NS moment decomposed}
    \\
    N_{mm'}^I &= \left[\bm F_I \sinh^2(\bm R) \bm F^\dagger_I\right]_{mm'},
    \label{eq:NI moment decomposed}
    \\
    M_{mm'}^{SI} &= \left[\bm F_S \cosh(\bm R)\sinh(\bm{R})  [\boldsymbol{F}_I]^{\mathrm{ T}}\right]_{mm'}.
    \label{eq:M moment decomposed}
\end{align}
The total number of pairs of signal and idler photons in the squeezed state{, $N_{SI}$, is obtained by summing over the diagonal entries of either the signal or idler moment in Eqs. \eqref{eq:NS moment decomposed} or \eqref{eq:NI moment decomposed} respectively,}
\begin{align}
    N_{SI} = \sum_{m} N^S_{mm}= \sum_{m} N^I_{mm} = \sum_{n=1}^{\ell} \sinh^2 (r_n).
    \label{eq:TotalNumber}
\end{align} 
An expression similar to Eq. \eqref{eq:efficiency} for the generation efficiency in the cw limit{,} and similar to Eq. \eqref{eq:LowGain} for the efficiency for the {pair regime}, can then be introduced here for the {LOS regime,}
\begin{equation}
    \mathcal{E}^{SI}_{\mathrm{pulse}} \equiv \frac{N_{SI}}{N_P}.
    \label{eq:PulsedEff}
\end{equation}
As we will see in our sample calculation, $ \mathcal{E}^{SI}_{\mathrm{pulse}}$ coincides with $ \mathcal{E}^{SI}_{\mathrm{pulse;pair regime}}$ for sufficiently low pulse energies and powers, as we would expect. To study the separability of the squeezed state in Eq. \eqref{eq:OutKetFinal4} we calculate its Schmidt number \cite{bib:Quesada2022,bib:Houde2023}
\begin{equation}
    K = \frac{\left(\sum_{n=1}^\ell \sinh^2 (r_n) \right)^2}{\sum_{n=1}^\ell \sinh^4 (r_n)},
    \label{eq:TotalSchmidtNumber}
\end{equation}
where $r_n$ are the entries of the diagonal matrix $\bm R$ in Eq. \eqref{eq:SVD_J}. {Within the approximations of this section, the Schmidt number in the {LOS regime} is unchanged from that in the {pair regime}, since the same pair generation operator from Eq. \eqref{eq:SuperA} appears in {Eq. \eqref{eq:OutKetFinalFirstOrder} from the pair regime} and {expression \eqref{eq:OutKetFinal2} from the LOS regime} for the output ket.  That is, within the approximations made here the joint spectral amplitude in the LOS regime is the same as the biphoton wave function in the pair regime.}

\section{\label{sec:NumResults}Sample Calculations}

We {now} present calculation{s} showing that structures yielding high generation efficiencies can be fabricated {from standard III-V semiconductors belonging to the $\bar{4}3m$ point group.} We simulate rings of Al$_{0.3}$Ga$_{0.7}$As (AlGaAs) and In$_{0.49}$Ga$_{0.51}$P (InGaP), fully encapsulated 
in {an} SiO$_2$ cladding \cite{CSOI_22}. The results are for rings of three different radii: 20 $\mu$m, 30 $\mu$m and 40 $\mu$m, and for various widths and heights of the {rings'} cross-section{s}. {A value of $\bar{\chi}^{(2)}=220$ pm/V is used for the nonlinear susceptibility, corresponding to the approximate value of the second-order nonlinear susceptibility for AlGaAs and InGaP \cite{yariv1984optical,bib:Adachi1985}}  

The pump is assumed to be polarized in the fundamental TM (TM0) mode around $775$ nm, and the signal and idler in the fundamental TE (TE0) mode around $1550$ nm: typical mode profiles are plotted in Fig. \ref{fig:ModeProfiles} {for an InGaP ring.} 
\begin{figure}[ht]
\centering
{
\includegraphics[scale=0.08]{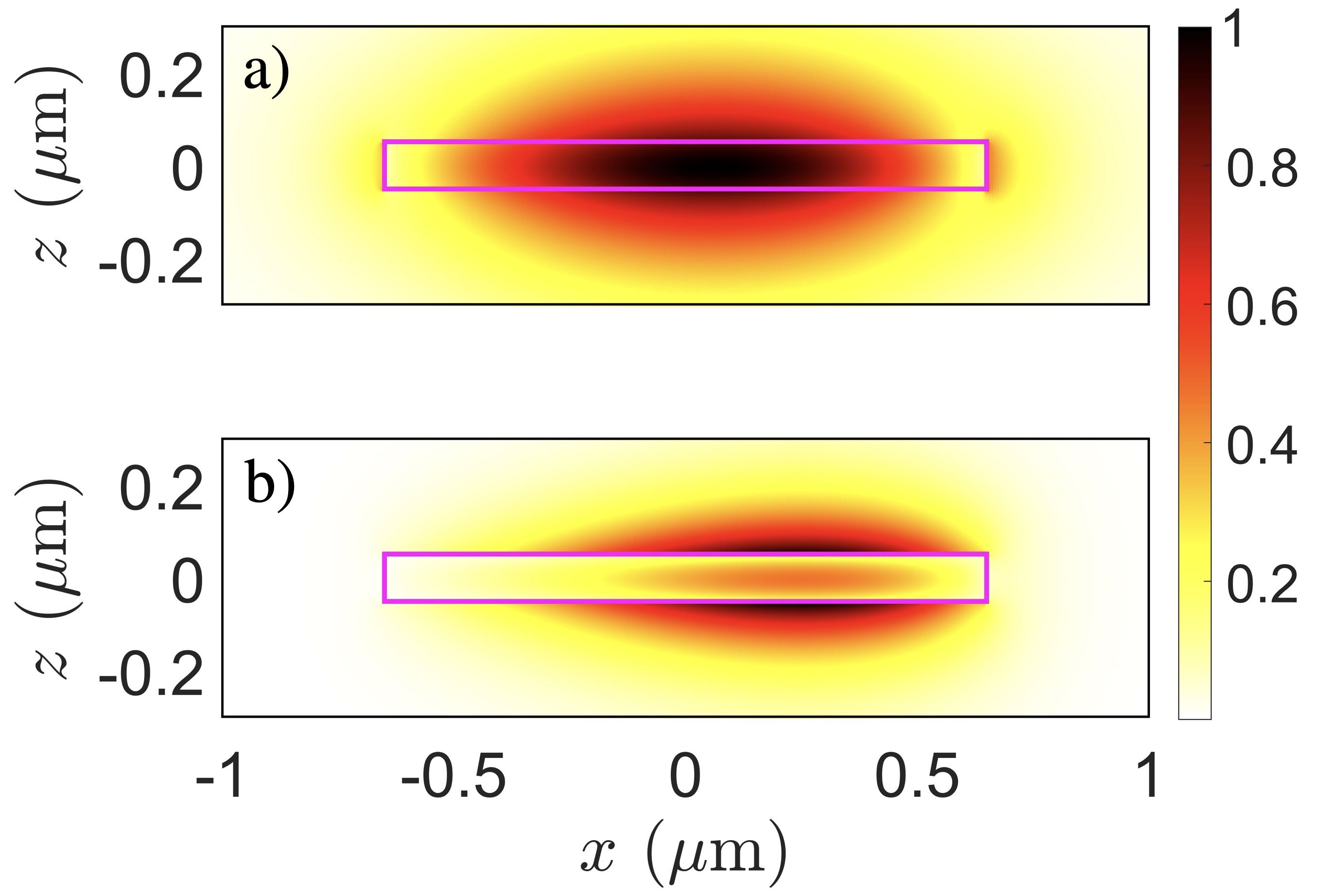}}
\centering
\caption{(a) {Electric-field $x$} component of the fundamental TE mode ($1550$ nm) for signal and idler, and (b)  {Electric-field $z$} component of the fundamental TM ($775$ nm) mode for the pump. Simulation done on Lumerical for InGaP with a height of $102$ nm, width of $1300$ nm and bend radius of $20\,\mu$m; {the centre of the ring is at negative $x$.} {The plotted fields are normalized by the maximum of the modulus squared of the total electric field.}}
\label{fig:ModeProfiles}
\end{figure}

{Consider first cw excitation at a frequency within the ring resonance at $\omega_P$, and the generation of signal and idler fields within the ring resonances centered at $\omega_S$ and $\omega_I$ respectively. This is possible only if the quasi-phase-matching condition from Eq. \eqref{eq:QuasiPhaseMatch_InText} is satisfied, 
which can be written as 
\begin{equation}
    \Delta\kappa=\mp 2/\mathcal{R},
    \label{eq:QPM}
\end{equation}
where 
\begin{equation}
\Delta\kappa=\frac{\omega_P n_{\mathrm{TM0}}(\omega_{P})}{c} - \frac{\omega_{S} n_{\mathrm{TE0}}(\omega_{S})}{c} - \frac{\omega_{I} n_{\mathrm{TE0}}(\omega_{I})}{c},   
\label{eq:UPMCond}
\end{equation} 
and $n_{\mathrm{TM0}}(\omega)$ and $n_{\mathrm{TE0}}(\omega)$ are the effective index functions for the pump, and the signal and idler, respectively; see Fig. \ref{fig:Indices}.}

As {a first} estimate {of ideal operating points, we put} $\omega_{S}\approx \omega_{I}\approx\frac{1}{2}\omega_P$, {and neglect the quasi-phase-matching term in Eq. \eqref{eq:QPM}, giving $\Delta\kappa=0$, or} 
\begin{equation}
n_{\mathrm{TM0}}(2\omega_S)= n_{\mathrm{TE0}}(\omega_S).
 \label{eq:ns}
\end{equation}
{In Fig. \ref{fig:Indices}, we plot $n_{\mathrm{TE0}}(2\pi c/\lambda_S)$ as a function of   $\lambda_S$, and $n_{\mathrm{TM0}}(2\pi c/\lambda_P)$ as a function of $2\lambda_P$, for different structures; the curves take into account the full dispersion of the modes for the TM0 mode around $775$ nm and the TE0 mode in the interval $[1500,1600]$ nm, and the intersections identify the instances where Eq. \eqref{eq:ns} is satisfied. That condition is satisfied for signal/idler wavelengths between $1520$ nm and $1540$ nm, for example, for {AlGaAs structures} with} cross-section dimensions (width$\times$height) of (1100$\times$105)nm and {InGaP structures with cross-section dimensions} (1300$\times$102)nm.

\begin{figure}[ht]
\centering{
\includegraphics[scale=0.13]{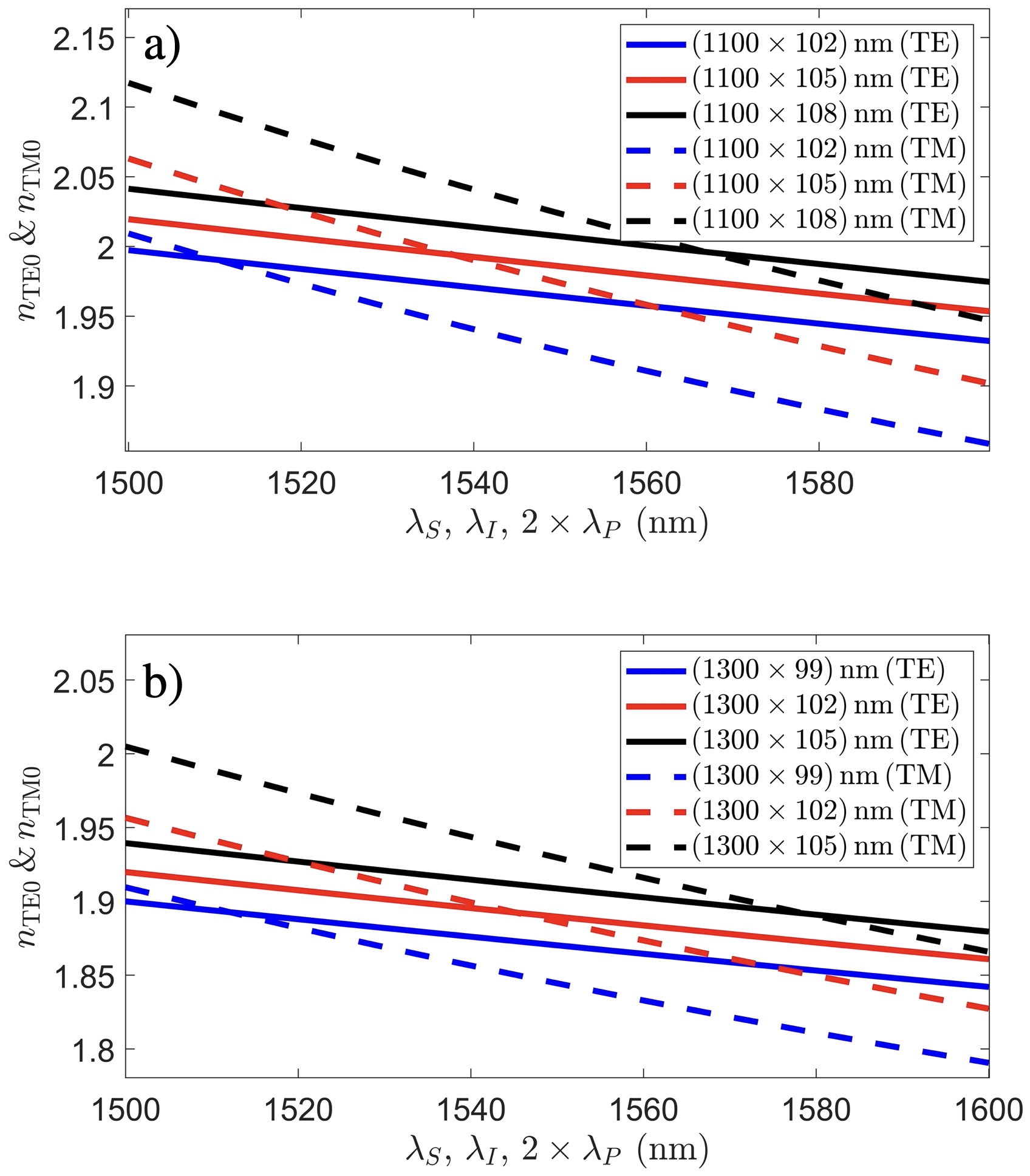}}
\centering
\caption{Effective indices of a) AlGaAs and b) InGaP as a function of wavelength for various types of structures with a radius of 30 $\mu$m. These plots show that the effective indices intersect around the wavelengths of interest.}
\label{fig:Indices}
\end{figure}

{We present calculations below that identify structures that can produce high generation rates, and the pump frequencies at which they do so. These calculations go beyond requiring the simple constraint from Eq. \eqref{eq:ns}, but we will see that condition indeed indicates the ranges of structures that are worth further investigation for signal/idler frequencies in the neighbourhood of the $\omega_S$ appearing there, both for cw excitation and for pulsed excitation.}

{In the calculations below we adopt realistic quality factors:} intrinsic quality factors of $1\times10^6$ at the signal and idler frequencies, and of $1\times10^5$ at the pump frequency \cite{akin2024ingap,Thiel24,PRXQuantum.2.010337}. {We consider a range of couplings, but at critical coupling the} simulated rings (with $\mathcal{R}=30\,\mu$m) have a finesse of around $\mathcal{F}_P^{\mathrm{ac}}\approx50$ and $\mathcal{F}_{S}^{\mathrm{ac}}=\mathcal{F}_{I}^{\mathrm{ac}}\approx1300$ (where $\mathcal{F}_u^\lambda = {v_u^{\lambda}}/{2\bar{\Gamma}_u \mathcal{R}}$ \cite{bib:Quesada2022}): {the field intensity in the ring is amplified by a factor of $\sim15$ for the pump frequency compared with the incident field, and the enhancement factor for the signal and idler is $\sim425$ \cite{bib:Quesada2022}}.

In the following sections, we present results for the efficiencies for both cw and pulsed excitation. 

\subsection{\label{subsec:RatesResults}cw excitation}

{We now go beyond the simple estimate of Eq. \eqref{eq:ns} in determining for what structures a large pair production rate can be achieved, and how large that rate can be.  We consider structures close to those identified by the intersections in Fig. \ref{fig:Indices}, and of course in evaluating the rate (Eq. \eqref{eq:Rate_Pairs}) need only consider signal and idler resonances for which the quasi-phase-matching condition from Eq. \eqref{eq:QPMuse} is satisfied. Each resulting $R^{SI}_{\lambda\lambda'}$ is optimized when 
\begin{equation}
    G(\omega_0)\equiv\left[ (\delta\omega)^2 + (\bar{\Gamma}_P)^2 \right]\left[(\Delta\omega)^2 +(\bar{\Gamma}_S+\bar{\Gamma}_I)^2 \right] 
    \label{eq:MinimizeRate}
\end{equation}
is minimized, where recall $\delta\omega=\omega_0-\omega_P$ and $\Delta\omega=\omega_0-\omega_S-\omega_I$.  In Fig. \ref{fig:Grids} (a-f) we plot the generation efficiencies $\mathcal{E}^{\mathrm{total}}_{\mathrm{ac,ac}}$ for both photons exiting the actual channel with $\lambda_S,\lambda_I\in[1500,1600]$ nm; {and in constructing the plots we have assumed} $\delta\omega=0$ and critical coupling. We refer to the range of structures investigated in each of the subplots {of Fig. \ref{fig:Grids}} as a ``family."} 
\begin{figure}[ht]
\centering{
\includegraphics[scale=0.103]{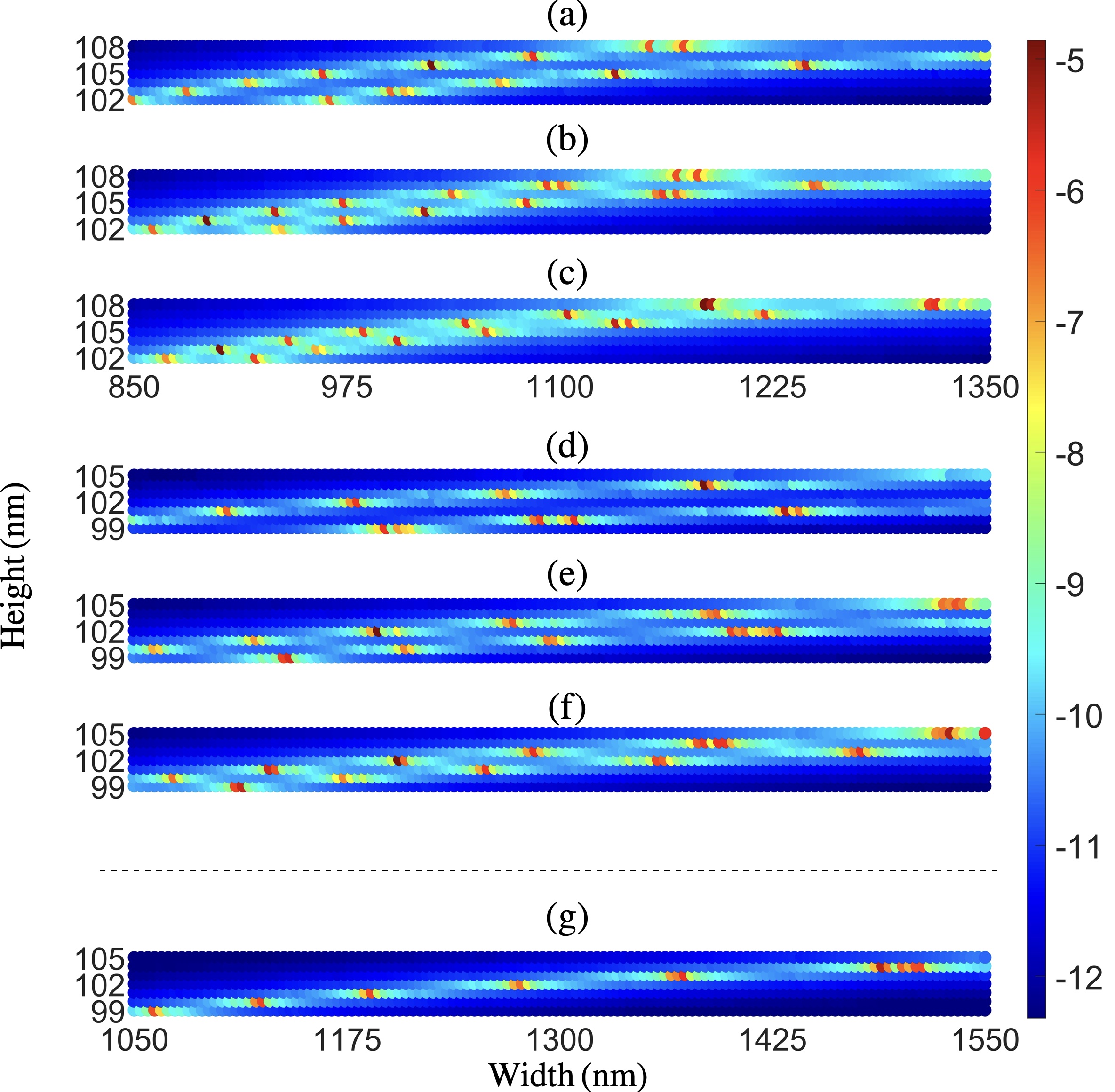}}
\centering
\caption{{Generation efficiencies. Grid of $\log_{10}(\mathcal{E}^{\mathrm{total}}_{\mathrm{ac,ac}})$} for varying widths and heights for an Al$_{0.3}$Ga$_{0.7}$As microring of (a) 20 $\mu$m, (b) 30 $\mu$m, (c) 40 $\mu$m radius and an In$_{0.49}$Ga$_{0.51}$P microring of (d) 20 $\mu$m, (e) 30 $\mu$m, (f, g) 40 $\mu$m radius. The shown efficiencies are those where both the signal and idler leave the actual channel. The plots are for critical coupling, $\bar{\chi}^{(2)}=220$ pm/V, $\lambda_S,\lambda_I\in[1500,1600]$ nm, $\lambda_P\approx775$ nm, $\delta\omega=0$, $Q^{\mathrm{(load)}}_S=Q^{\mathrm{(load)}}_I=5\times10^5$, $Q^{\mathrm{(load)}}_P=5\times10^4$. Plots a)--f) are for the physical {quasi-phased matched (QPM)} cases ($\Delta\kappa\pm2/\mathcal{R}=0$), and plot g) is for the artificial {usual phase matched (UPM)} case ($\Delta\kappa=0$), to be compared with figure f).}
\label{fig:Grids}
\end{figure}

The structures yielding high generation efficiencies are given by the red spots in Fig. \ref{fig:Grids} {(a-f)}, {which we refer to as ``hotspots,"} and {at these} in general only one set of quasi-phase-matched ring resonances contributes significantly to the total rate. The highest efficiencies from each {family of structures investigated in} Fig. \ref{fig:Grids} are tabulated in Tab. \ref{tab:TableMax}. For excitation at wavelength $\lambda_0\approx775$ nm, {the} optimized generation efficiencies range from $0.67-1.39\times10^{-5}$. These efficiencies are in accord with {recent} experimental results \cite{Zhao:22}. 

Allowing the pump frequency $\omega_0$ to differ from the centre frequency $\omega_P$ of the pump resonance leads to at most only minor increases in the efficiencies, and to a change in the structure within each family for which the peak efficiency is obtained.

To examine the effects of varying the coupling between the ring and the channel, we assume that the escape efficiencies (Eq. \eqref{eq:escape efficiencies}) of the signal and idler are the same, $\eta_{S}^{\mathrm{ac}}=\eta_{I}^{\mathrm{ac}}$, and in Fig. \ref{fig:EscEff2} we plot the ratio {\textbf{R}} of the generation rate into the actual channel to its value at critical coupling,  
\begin{equation}
\textbf{R}=\frac{R^{SI}_{\mathrm{ac,ac}}(\eta_S^{\mathrm{ac}}=\eta_I^{\mathrm{ac}})}{R^{SI}_{\mathrm{ac,ac}}(\eta_S^{\mathrm{ac}}=\eta_I^{\mathrm{ac}}=1/2)}.
\label{eq:generation_ratio}
\end{equation}
This ratio peaks at $\eta_{S}^{\mathrm{ac}}=\eta_{I}^{\mathrm{ac}}=2/3$, where it takes the value 32/27. If we envision a heralding experiment where the signal photon is used to herald the idler photon, the heralding efficiency, $\mathcal{E}_\mathrm{herald}$, is given by 
\begin{equation}    \mathcal{E}_\mathrm{herald}=\frac{R^{SI}_{\mathrm{ac,ac}}}{R^{SI}_{\mathrm{ac,ac}}+R^{SI}_{\mathrm{ac,ph}}}  .
\end{equation}
If we again put $\eta_{S}^{\mathrm{ac}}=\eta_{I}^{\mathrm{ac}}\equiv\eta^{\mathrm{ac}}$, then 
the heralding efficiency is just equal to  the escape efficiency of the actual channel $\mathcal{E}_\mathrm{herald}=\eta^{\mathrm{ac}}$, which we indicate in Fig. \ref{fig:EscEff2}. 

\begin{figure}[ht]
\centering{
\includegraphics[scale=0.0395]{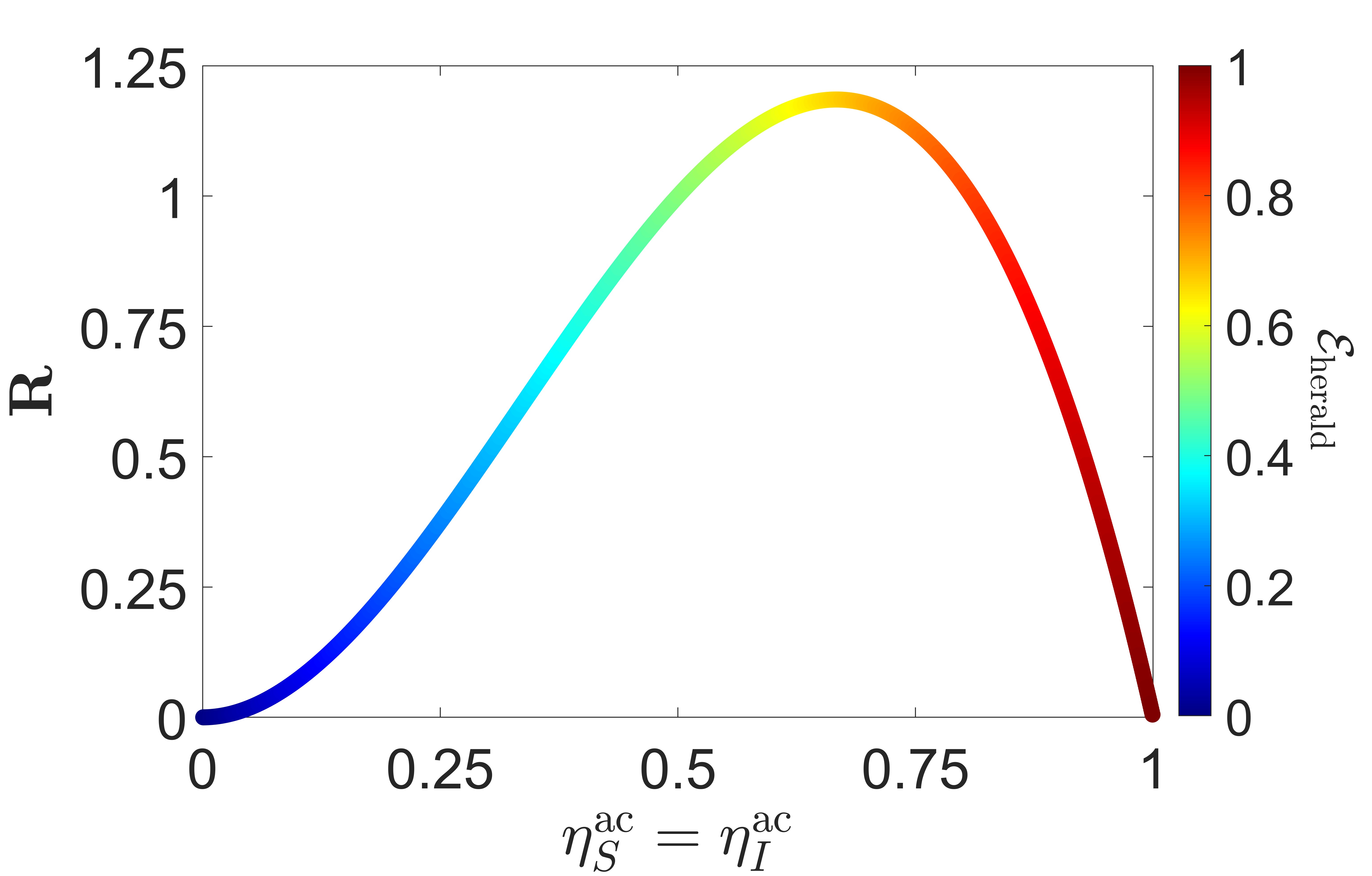}}
\centering
\caption{{The ratio of generation efficiency {for} both photons exit{ing} the actual waveguide to {the corresponding efficiency at} critical coupling, and {the} heralding efficiency for the respective generation efficiencies. The maximal generation efficiency {occurs at} escape efficiencies of $\eta_S^{\mathrm{ac}}=\eta_I^{\mathrm{ac}}=2/3$, {where the heralding efficiency is 2/3.}}}
\label{fig:EscEff2}
\end{figure}

We note that having to satisfy {one of} the quasi-phase-matching {(QPM)} condition{s} $\Delta\kappa\pm2/\mathcal{R}=0$ instead of {the usual} phase matching {(UPM)} condition $\Delta\kappa=0$ {(see Eqs. \eqref{eq:UPMCond}-\eqref{eq:ns} and the discussion following)} does not necessarily {lead to an} increase {in} the {generation efficiency} for a given {family of structures of the type considered here.} 

{In Fig. \ref{fig:Grids} (g) we plot the generation efficiencies that would result for the structures used in Fig. \ref{fig:Grids} (f) if we arbitrarily imposed the UPM condition instead of the QPM condition. The maximum generation efficiencies in Fig. \ref{fig:Grids} (f) and Fig. \ref{fig:Grids} (g) are comparable, although they occur for different structures.  And for each family of structures there are twice as many {``hotspots"} in the actual scenario of Fig. \ref{fig:Grids} (f) as there are in the artificial scenario of Fig. \ref{fig:Grids} (g), for there are twice as many QPM conditions as there are UPM conditions.}  

We give a quantitative example: for the InGaP ring with radius of 40 microns {(}(f) and {(g)} from Fig. \ref{fig:Grids}), the single UPM hotspot {is a structure with} a height of $102$ nm, {a} width of $1278$ nm, and an efficiency of $0.28\times10^{-6}$; the minus and plus QPM hotspots for the same $102$ nm height have widths of $1206$ nm and $1358$ nm {respectively,} {and have} efficiencies of $7.72\times10^{-6}$ and $0.39\times10^{-6}$ respectively. The mode numbers of the resonances contributing the most to the rates for the UPM condition are $m_P=612$, $m_S=314$, and $m_I=298$ (with wavelengths $\lambda_P=774.87$ nm, $\lambda_S=1522.93$ nm, and $\lambda_I=1577.52$ nm). The mode numbers of the resonances contributing the most to the rates for the minus and plus QPM hotspots are, respectively: $m_P=611$, $m_S=315$, and $m_I=294$ (with wavelengths $\lambda_P=774.96$ nm, $\lambda_S=1514.96$ nm, and $\lambda_I=1586.52$ nm), and $m_P=613$, $m_S=319$, $m_I=296$ (with wavelengths $\lambda_P=774.78$ nm, $\lambda_S=1511.24$ nm, and $\lambda_I=1589.86$ nm).

\begin{table*}[tp]
\centering
{
\begin{tabular}{c||c|c|c|c|c|c|c}

Material                & Fig. \ref{fig:Grids} & $\mathcal{R}$ ($\mu$m)  & Dim. (nm$^2$)  & $\mathcal{E}^{\mathrm{total}}_{\mathrm{ac,ac}}$ & $R^{\mathrm{total}}_{\mathrm{ac,ac}}$ (MHz) ($P_P=1$ $\mu$W) & Effective area $A_{\mathrm{eff}}$ ($\mu$m$^2$) [$\pm$] & $\Delta\omega/2\pi$ (MHz)\\ \Xhline{3\arrayrulewidth}
\multirow{3}{*}{AlGaAs} & a)                   & 20                      & $1026 \times 106$    & $1.04\times10^{-5}$ & $40.5$ &  $0.519$  [$-$]  & $-340$   \\ \cline{2-8} 
                        & b)                   & 30                      & $894  \times 103$    & $0.67\times10^{-5}$ & $26.3$ & $0.563$  [$-$]  & $-324$   \\ \cline{2-8} 
                        & c)                   & 40                      & $1186 \times 108$    & $0.88\times10^{-5}$ & $34.4$ & $0.537$  [$-$]  & $-27$   \\ \hline
\multirow{3}{*}{InGaP}  & d)                   & 20                      & $1386 \times 104$    & $1.39\times10^{-5}$ & $54.4$ & $0.568$  [$-$]  & $-173$   \\ \cline{2-8} 
                        & e)                   & 30                      & $1194 \times 102$    & $1.17\times10^{-5}$ & $45.6$ & $0.557$  [$-$]  & $-17$  \\ \cline{2-8} 
                        & f)                   & 40                      & $1206 \times 102$    & $0.77\times10^{-5}$ & $30.1$ & $0.567$  [$-$]  & $113$   
\end{tabular}}
     \caption{Largest generation efficiencies of the different {families of structures in the} subfigures of Fig. \ref{fig:Grids}, with the respective effective areas and the values of $\Delta\omega$ for the set of resonances that contributes the most to the rates. The $\pm$ in square brackets {indicates} which quasi-phase-matching condition {is} met for the resonances that contribute the most to the generation rate. We have taken $\bar{n}_P=\bar{n}_S=\bar{n}_I=\bar{n}=3.5$, and  $\bar{v}_P=\bar{v}_S=\bar{v}_I=c/\bar{n}$.}
     \label{tab:TableMax}
\end{table*}

\begin{figure}[t]
\centering
{
\includegraphics[scale=0.068]{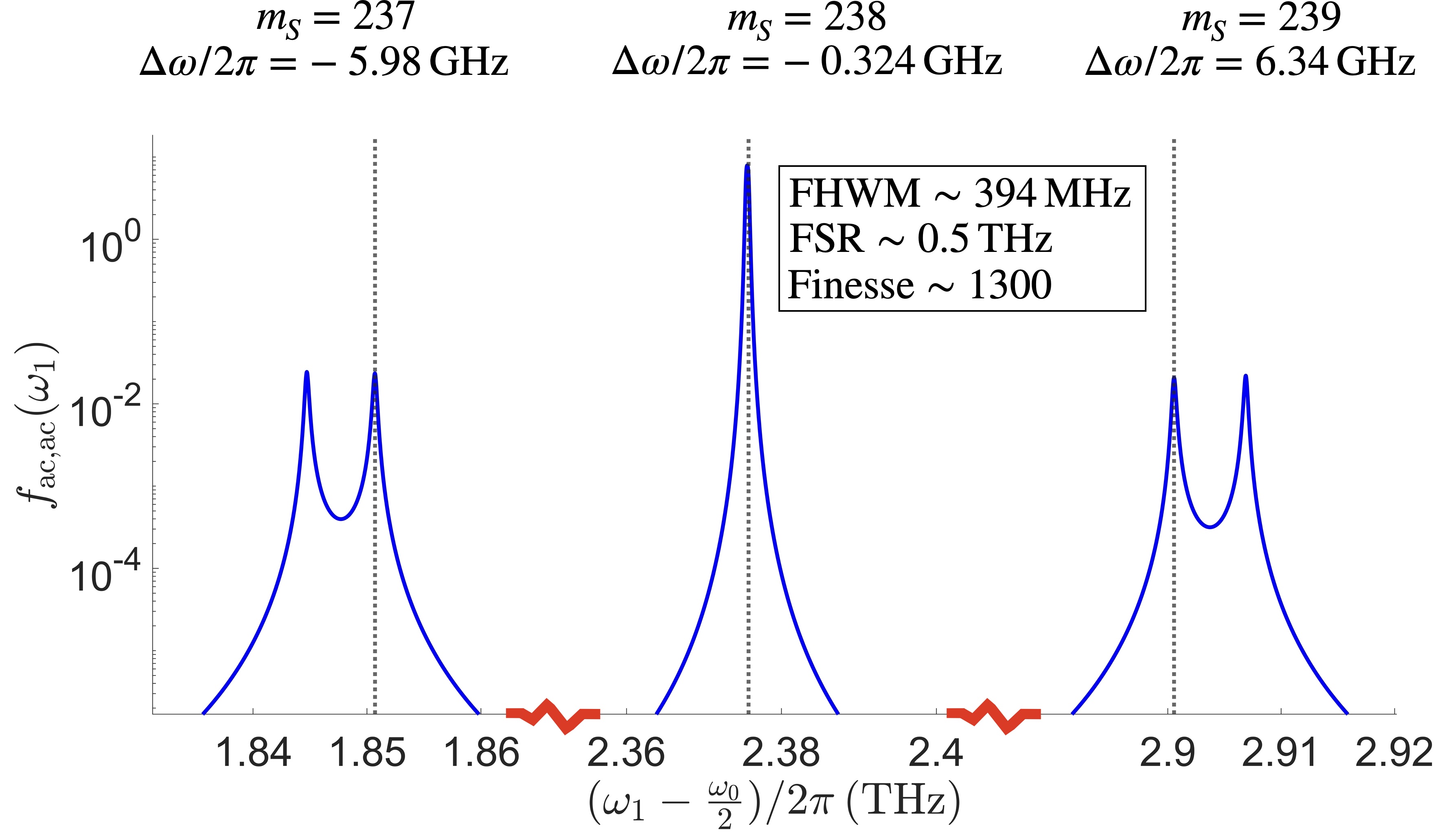}}
\centering
\caption{Spectrum of the signal photons exiting the actual waveguide for the optimized AlGaAs microring with $\mathcal{R}=30\,\mu$m, cross-sectional dimensions of {$894\,\mathrm{nm}\times103$ nm} (b) from Tab. \ref{tab:TableMax}). Critical coupling is assumed, $\bar{\chi}^{(2)}=220$ pm/V, $\lambda_S\in[1500\,\mathrm{nm},2\times \lambda_P]$ nm, $\lambda_P=775.23$ nm, $\delta\omega=0$, $Q^{\mathrm{(load)}}_S=Q^{\mathrm{(load)}}_I=5\times10^5$, $Q^{\mathrm{(load)}}_P=5\times10^4$. The dashed lines represent the centre frequencies of the different signal resonance bins $\omega_S$. The idler frequency is given by $\omega_2=\omega_0-\omega_1$, the pump mode number is $m_P=469$, and the idler mode numbers are given by $m_I=m_P-m_S\pm2$.}
\label{fig:RateDist}
\end{figure}

{We can write Eq. \eqref{eq:Rworkout} for the generation rate of signal and idler photons both in the actual channel as}
\begin{equation}
    R_{\mathrm{ac,ac}}^{\mathrm{total}} = \int_{\omega_0/2}^{2\pi c / 1500\,\mathrm{nm}} \mathrm{d}\omega_1 f_{\mathrm{ac,ac}}(\omega_1), 
    \label{Eq:Spectrum}
\end{equation}
where 
\begin{equation}
    f_{\mathrm{ac,ac}}(\omega_1) = \frac{2 \pi}{\hbar^2}  \frac{|\mathsf{M}_{\mathrm{ac,ac}} (k_1(\omega_1),k_2(\omega_0-\omega_1))|^2}{v_S^{\mathrm{ac}} v_I^{\mathrm{ac}}},
    \label{eq:Spectrum_M}
\end{equation}
and 
\begin{equation}
\omega_i\equiv\omega_{uk_i}^{\lambda}=\omega_u + v_u^{\lambda}(k_i-K_u^{\lambda}).
\label{eq:NewOmega}
\end{equation}
{Here we have used Eq. \eqref{eq:basic_rate} and Eq. \eqref{eq:Rate_PairsDegen}  for $R^{SI}_{\mathrm{ac,ac}}$ and $\hat{R}^{SS}_{\mathrm{ac,ac}}$ respectively, changed variables, and integrated over the idler} frequency $\omega_{Ik_2}^{\mathrm{ac}}$ with the Dirac delta function. {Thus} $f_{\mathrm{ac,ac}}(\omega_1)$ {can be identified as the spectrum of the signal photon when both photons exit the actual channel, and in Fig. \ref{fig:RateDist} we show it for} the optimized AlGaAs ring with $\mathcal{R}=30\,\mu$m, {restricting ourselves to three resonances, the centre one being the one that contributes the most to the rate;} the other resonances contribute very little to the rate. We see that the spectrum is highly peaked for one resonance, which dominates the rate, {and} this is true in general for the hotspots we see in Fig. \ref{fig:Grids}. The \emph{{double peaks}} in the spectrum come from the two field-enhancement factors that appear in the spectrum $f_{\mathrm{ac,ac}}(\omega_1)$: these two field-enhancement factors are Lorentzians peaked at {$\omega_1=\omega_S$ and $\omega_1=\omega_0-\omega_I$ respectively. The first of these corresponds to the signal being on resonance, and the second to the idler being on resonance, $\omega_0-\omega_1=\omega_I$.} If the frequency bin matching condition {were met exactly,} i.e. $\omega_S=\omega_0-\omega_I$ {and $\Delta\omega=0$,} the two Lorentzians would peak at the same frequency {$\omega_1$, and} only a single peak per resonance would appear in the spectrum; {usually this does not occur, and two peaks are typically associated with each resonance, although in some instances they are not resolved.}

As can be seen from Fig. \ref{fig:Grids}, the generation efficiencies, {which depend on the effective indices of the modes,} are much more sensitive {to variations in the heights of the guides than to variations in their widths, as might be expected from the confinement of the modes in the waveguide as shown in Fig. \ref{fig:ModeProfiles}. } {Thus the predicted peak efficiencies are sensitive to any discrepancies between the assumed bulk indices and the actual values in integrated optical structures, and indeed even to the precision of numerical calculations.} However, {even with generous assumption of uncertainties,} we have found hotspots close to the values of those plotted in Fig. \ref{fig:Grids}; {and there are very generally} large diagonal bands like those shown in Fig. \ref{fig:Grids} that enclose a multitude of hotspots and high efficiencies. {Thus it should be possible to identify structures that yield high generation efficiencies regardless of these uncertainties.} 

We stress that the rate equation and conversion efficiencies discussed thus far are in the weak pumping limit; at large enough pump powers the Fermi's golden rule calculation presented here will break down. 
This is more easily explored, and may be more relevant, in the scenario of pump pulse excitation, which we consider in the next section.

\subsection{\label{subsec:BWFResults}Pulsed excitation}
\begin{figure}[ht]
\centering
{
\includegraphics[scale=0.12]{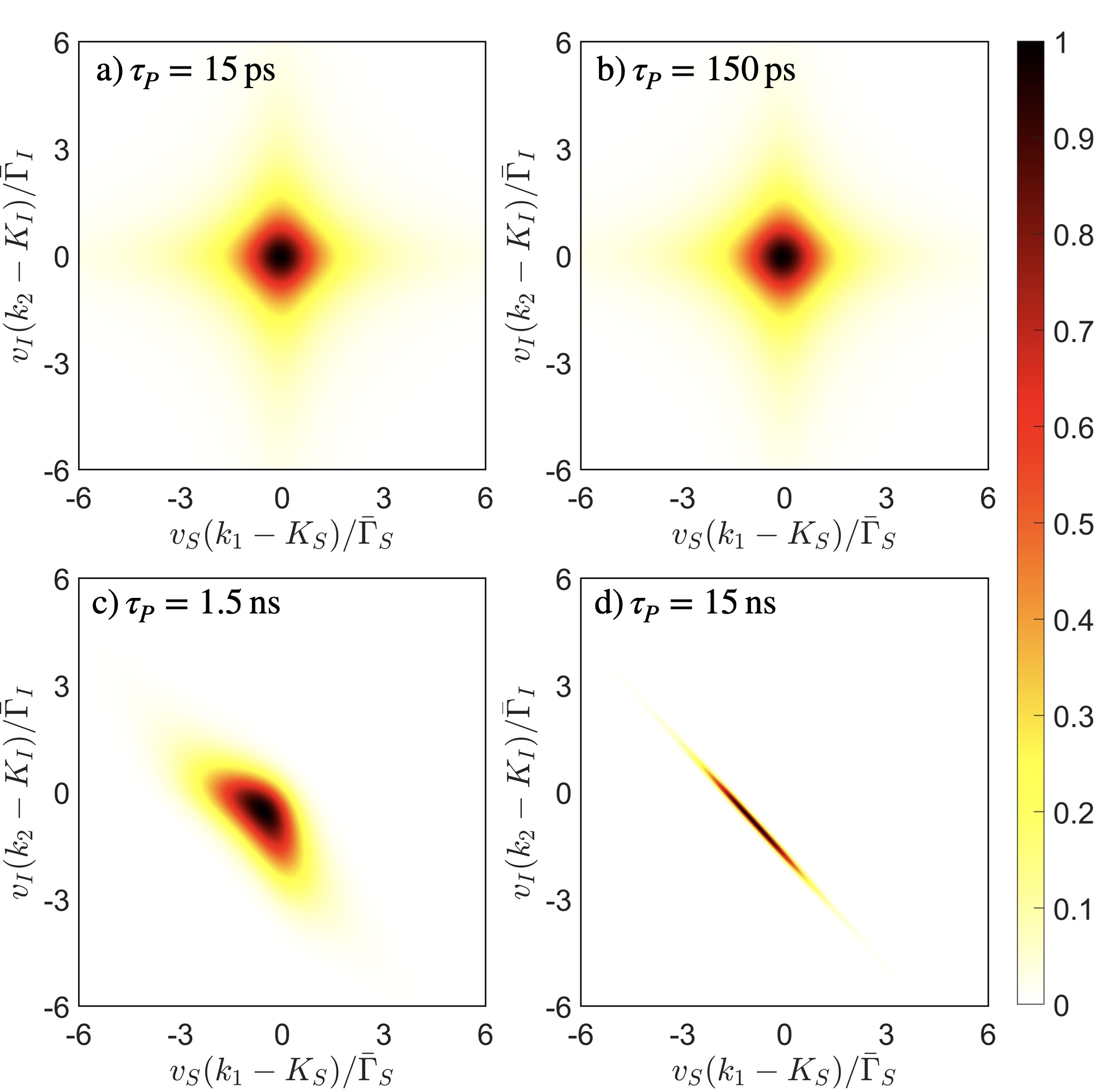}}
\centering
\caption{Joint spectral amplitudes {(we plot $\big(|\varphi^{\mathrm{II}}(k_1,k_2)|^2/\mathrm{max}\left[|\varphi^{\mathrm{II}}(k_1,k_2)|^2\right]\big)$)} {from a ring of radius $\mathcal{R}=30\,\mu$m, using resonances $\lambda_P=775.23\,\mathrm{nm}$, $\lambda_S=1531.64\,\mathrm{nm}$, $\lambda_I=1569.75\,\mathrm{nm}$, and with $\Delta\omega/2\pi=-324$ MHz and $\delta\omega=0$.} {The} pump pulses {are Gaussian with various FWHM $\tau_P$ as indicated, and} peak power{s of} $P_{P}=10\,\mu$W; they have energies $E_P=$ (a) $0.16$ fJ, (b) $1.6$ fJ, (c) $16$ fJ, (d) $0.16$ pJ. {The Schmidt numbers of the resulting joint spectral amplitudes are} $K=$ (a) $1.00031$, (b) $1.0028$, (c) $1.31$, (d) $8.38$. The total number of signal photons in {the} outgoing state{s are} $N_{SI}=$ (a) $6.50\times10^{-4}$, (b) $0.0619$, (c) $2.49$, (d) $26.58$, and $|\beta|^2=$ (a) $6.50\times10^{-4}$, (b) $0.0607$, (c) $1.70$, (d) $16.43$.}
\label{fig:BiPhotonWFs2}
\end{figure}

Due to the similarities of the equations for cw and pulsed excitation, {we can expect that} the optimized structures yielding high efficiencies for cw {excitation} also yield high efficiencies for pulsed {excitation.} {So we focus on} one of our optimized rings from Tab. \ref{tab:TableMax}: the AlGaAs ring of radius $\mathcal{R}=30\,\mu$m and cross-sectional dimensions of {$894\,\mathrm{nm}\times103$ nm}. {We take the coupling to be critical, and consider Gaussian pulses, 
\begin{equation}
    \phi(k) = \left[\frac{\tau_P^2(v_P^{\mathrm{ac}})^2}{8\ln(2)\pi}\right]^{\frac{1}{4}}\exp\left\{\frac{-\tau_P^2 (v_P^{\mathrm{ac}})^2\left[k-K_P^{\mathrm{ac}}-\frac{\delta\omega}{v_P^{\mathrm{ac}}}\right]^2}{16\ln(2)}\right\},
    \label{eq:NormalizedPulseInK}
\end{equation}
with temporal full width half maximum values $\tau_P$ ranging from $15$ ps to $15$ ns, always with a peak power of $10$ $\mu$W; the pump pulse energies then range from $0.16$ fJ to $0.16$ pJ. The set of resonances identified by the} mode numbers $m_P=469$, $m_S=238$ and $m_I=229$, {with} centre wavelengths $\lambda_P=775.23\,\mathrm{nm}$, $\lambda_S=1531.64\,\mathrm{nm}$, and $\lambda_I=1569.75\,\mathrm{nm}$, {contributes most to the generation of photon pairs. Here} the plus quasi-phase-matching condition is satisfied, and $\Delta\omega/2\pi=-324$ MHz: this is comparable to the sum of the linewidths of the signal and idler $(\bar{\Gamma}_S+\bar{\Gamma}_I)/2\pi=387\,\mathrm{MHz}$. 

The {joint spectral amplitudes} are shown in Fig. \ref{fig:BiPhotonWFs2}. {We see that for realistic pump pulses such a structure can be used to} generate pairs that have both low ($1.00031$), and high ($8.38$) Schmidt numbers, {depending on the duration of the pump pulse.} By varying the coupling {even higher and lower Schmidt numbers can be obtained.} {For the} shorter and lower energy pulses ({Fig. \ref{fig:BiPhotonWFs2} (a) and (b)}), the number of generated signal photons $N_{SI}$ calculated from Eq. \eqref{eq:TotalNumber} {essentially equals} $|\beta|^2$, {as expected} in the {pair regime} from Eq. \eqref{eq:PairsInPulse}. {But for longer and higher energy pulses (Fig. \ref{fig:BiPhotonWFs2} (c) and (d)) we see that the {approximation {of} the pair regime} fails, as $N_{SI}$ becomes significantly larger than $|\beta|^2$.}

\begin{figure}[t]
\centering
{
\includegraphics[scale=0.059]{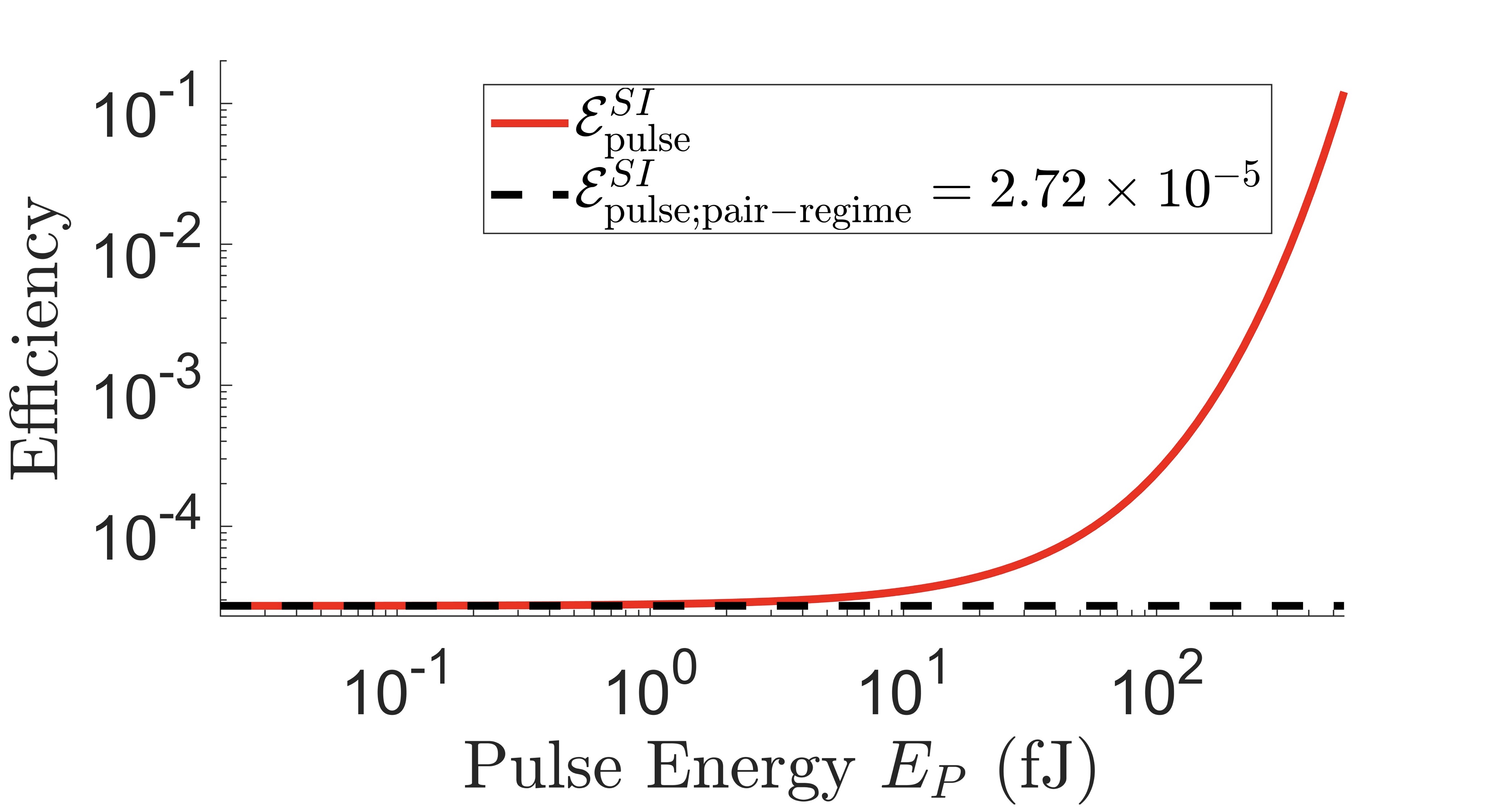}}
\centering
\caption{Scaling of the generation efficiency for varying pulse energies for the optimized AlGaAs ring of radius $\mathcal{R}=30\,\mu$m, where $\lambda_P=775.23\,\mathrm{nm}$, $\lambda_S=1531.64 \,\mathrm{nm}$, $\lambda_I=1569.75 \,\mathrm{nm}$ and $\Delta\omega/2\pi=-324$ MHz. The pump pulse duration is fixed: $\tau_P=1.5$ ns. The {LOS} {approximation} (solid red line) diverges from the {approximation from the pair regime} (dashed black line) for pulse energies of around 1–10 fJ, corresponding to peak powers of 0.63–6.3 $\mu$W.}
\label{fig:PreDepletion}
\end{figure}

To investigate {this further, with the same} optimized 30 $\mu$m AlGaAs ring from Tab. \ref{tab:TableMax} {we consider excitation by pump pulses all of duration} $\tau_P=1.5$ ns, {but with varying pulse energies,} (and hence peak powers). {In} Fig. \ref{fig:PreDepletion} {we plot the predicted efficiencies from our {approximation for both the pair regime} and {LOS regime}, and we see that for} relatively low and realistic pulse energies {not only does the {pair regime} approximation fails, but the {LOS} approximation predicts efficiencies so large that pump depletion -- which has been neglected here -- would have to be taken into account to determine a reliable estimate.} Future work will include both time-ordering corrections, {which have been neglected in the simple {LOS} approximation presented here,} and pump depletion.

\section{\label{sec:Conclusion}Conclusion and Future Work}
{We have derived general formulas for the generation rate of photon pairs, {and the generation of squeezed light in the lowest-order squeezing (LOS) regime,} from the process of SPDC in ring resonator structures. {Since our approach treats the scattering losses quantum mechanically, we can also analyze the} statistics of the scattered photons. {We have focused on microring resonators made of} III-V {semiconductors,} point-coupled to a waveguide, and {our approach} can easily be extended to multiple input waveguides, and to microrings made from different materials. 

{With the integrated photonic structures we have considered, where modal dispersion is significant, the quasi-phase-matching associated with the tensor nature of the nonlinearity does not play an essential role in obtaining high efficiencies, although it does affect what structures will lead to the highest efficiencies.} From our sample calculation{s}, we obtain optimized structures yielding high generation efficiencies {of $2.72\times 10^{-5}$} in  the {pair regime}, {and we have detailed how both the rate at which pairs of generated photons exit the channel without loss, and the heralding efficiencies, depend on the coupling between the channel and the resonator.} 

{In the LOS regime we find efficiencies} {in excess of  $1\times 10^{-2}$} {for realistic structures, and indeed such structures should lead to elevated levels of squeezing and pump depletion beyond the LOS regime.}   

{In future work we will extend} this calculation {to treat the higher levels of squeezing that are possible, and include third-order processes such as self- and cross-phase modulation that will affect the dynamics of the squeezing.} 
{In the high-squeezing regime our calculations predict significant pump depletion. As pointed out earlier \cite{yanagimoto22}, in this regime the entanglement between the pump and the squeezed state leads to the generation of non-Gaussian states. In future work we will extend our calculation to include the non-Gaussian dynamics between the pump, signal, and idler fields. As we have shown, high efficiency microring resonators made from III-V semiconductor materials are a promising platform for investigating non-Gaussian states generated via nonlinear interactions. Furthermore, non-Gaussian states generated this way offer a potentially more efficient approach to introducing non-Gaussianity compared with probabilistic approaches based on heralding and photon-number-resolving (PNR) detection \cite{Bourassa2021}.}

\section*{Acknowledgements}
{Those of us at the University of Toronto would like to thank the Natural Sciences and Engineering Research Council of Canada for financial support. S. E. F. acknowledges support from a Walter C. Sumner Memorial Fellowship. J. E. S. and C. V. acknowledge support from the Horizon-Europe research and innovation program under grant agreement ID: 101070168 (HYPERSPACE).} UCSB acknowledges support from the NSF (CAREER-2045246), AFOSR (FA9550-23-1-0525), and the UC Santa Barbara NSF Quantum Foundry funded via the Q-AMASE-i program under award DMR-1906325. M. L. acknowledges PNRR MUR project ``National Quantum Science and Technology Institute" NQSTI (Grant No. PE0000023) .

\bibliography{apssamP}

\appendix
\section{laboratoryand ring frames}
\label{app:RefFrames}
With $\chi^{(2)}(\boldsymbol{r})$ the second-order nonlinear susceptibility of the material at $\boldsymbol{r}$, {take} $\chi^{(2)}_{ijk}{(\boldsymbol{r})}$ and $\chi^{(2)}_{i'j'k'}(\boldsymbol{r})$ {to be} the components of the tensor in the ring frame and the laboratoryframe respectively, with $i,j,k\in\{x,y,z\}$ and $i',j',k'\in\{x',y',z'\}$ {(see Fig. \ref{fig:Struct})}. For a zincblende structure grown in the $\hat z$ direction, the only non-zero components of ${\chi^{(2)}}{(\boldsymbol{r})}$ for $\boldsymbol{r}\in$ ring in the laboratoryframe are:
\begin{equation}
    \chi^{(2)}_{[x'y'z']}(\boldsymbol{r})=\bar{\chi}^{(2)},
    \label{ChiCalc[XYZ]_apP}
\end{equation}
where $\bar{\chi}^{(2)}$ is a constant and the value of the second order nonlinear susceptibility in the material, {and $[x'y'z']$ indicates any distinct permutation of $\{x',y',z'\}$}.

Let $\boldsymbol{M}_\phi$ be the {matrix that transforms the coordinates from the laboratoryframe to the ring frame, by rotation of {an} angle $\phi$: }
\begin{equation}
\boldsymbol{M}_\phi=\begin{bmatrix}\cos\phi&\sin\phi&0\\-\sin\phi&\cos\phi&0\\0&0&1\end{bmatrix}.
    \label{ChiCalcM}
\end{equation}
{For a vector $\boldsymbol{v}$ we have}
{\begin{equation}
\begin{bmatrix}v_x\\v_y\\v_z\end{bmatrix}=\boldsymbol{M}_\phi\begin{bmatrix}v_{x'}\\v_{y'}\\v_{z'}\end{bmatrix},
\end{equation}} {and since} ${\chi^{(2)}}{(\boldsymbol{r})}$ is a rank-three tensor
\begin{equation}
    \chi^{(2)}_{ijk}{(\boldsymbol{r})}=\boldsymbol{M}_\phi^{{i}i'}\boldsymbol{M}_\phi^{{j}j'}\boldsymbol{M}_\phi^{{k}k'}\chi^{(2)}_{i'j'k'}{(\boldsymbol{r})}.
    \label{ChiCalcMIiMJjMKk}
\end{equation}
{From} Eqs. \eqref{ChiCalc[XYZ]_apP}, \eqref{ChiCalcM} and \eqref{ChiCalcMIiMJjMKk} the only non-zero components of ${\chi^{(2)}}{(\boldsymbol{r})}$ in the ring frame are {found to be those of Eq. \eqref{eq:Chi2SinCos_InText}.}

\section{\label{app:DegenerateCase}Degenerate SPDC}

{In this Appendix we} treat degenerate SPDC, where both of the generated photons are within one frequency bin labelled by the centre frequency $\omega_S$. For variables that differ from those for nondegenerate SPDC, we will include a ``hat" for the degenerate case. The Hamiltonian that describes the degenerate SPDC interaction is
\begin{equation}
\begin{split}
    \hat{H}^{\mathrm{SPDC}}_{NL} = -\frac{1}{\epsilon_0} \int \Gamma^{(2)}_{i'j'k'}(\boldsymbol{r}) & \left[{\mathsf{D}}_{S}^{\mathrm{out}(i')}(\boldsymbol{r})\right]^\dagger \left[ {\mathsf{D}}_{S}^{\mathrm{out}(j')}(\boldsymbol{r})\right]^\dagger \\
    & \times {\mathsf{D}}_{P}^{\mathrm{in}(k')}(\boldsymbol{r}) \mathrm{d} \boldsymbol{r}
    + \mathrm{H.c.}.
\end{split}
    \label{eq:H_NL_SPDC_NewSumDegen}
\end{equation} 
We rewrite the Hamiltonian {in a form} similar to Eq. \eqref{eq:H_NL}:
\begin{equation}
\begin{split}
    \hat{H}^{\mathrm{SPDC}}_{NL} & = - \sum_{\lambda\lambda'} \int  \mathrm{d} k_1 \mathrm{d} k_2 \mathrm{d} k_3 \hat{\mathsf{K}}_{\lambda\lambda'}({k}_1,{k}_2,{k}_3) \\
    & \times \left[a_{S}^{\mathrm{out},\lambda}(k_1)\right]^\dagger \left[ a_{S}^{\mathrm{out},\lambda'}(k_2)\right]^\dagger a_{P}^{\mathrm{in},\mathrm{ac}}(k_3) + \mathrm{H.c.},
    \label{eq:H_NLDegen}
    \end{split}
\end{equation}
where {here}
\begin{equation}
\begin{split}
    \hat{\mathsf{K}}_{\lambda\lambda'}(k_1,k_2,k_3) = & \sqrt{\frac{\hbar \omega_P}{4 \pi} \frac{\hbar \omega_{S}}{4 \pi} \frac{\hbar \omega_{S}}{4 \pi}}   \left[ F_{S+}^{\lambda}(k_1) F_{S+}^{\lambda'}(k_2) \right]^* \\
    & \times F_{P-}^{\mathrm{ac}}(k_3) \times \hat{\bar{\mathsf{K}}}_{SSP},
    \label{eq:K_ExpressionDegen}
    \end{split}
\end{equation}
and
\begin{equation}
\begin{split}
    \hat{\bar{\mathsf{K}}}_{SSP}  &=  \epsilon_0 \int \chi^{(2)}_{i'j'k'}(\boldsymbol{r}) [ {\mathsf{e}}_{S}^{i'}(\boldsymbol{r}) {\mathsf{e}}_{S}^{j'}(\boldsymbol{r}) ]^* {\mathsf{e}}_{P}^{k'}(\boldsymbol{r}) \\
    & \,\,\,\,\,\,\,\,\,\,\,\,\,\,\,\,\,\,\,\,\,\,\,\,\,\, \times e^{i(\kappa_P - \kappa_{S} - \kappa_{S})\zeta} \mathrm{d} \boldsymbol{r}_\perp  \mathrm{d} \zeta\\ 
    & =\frac{1}{2}\epsilon_0 \bar{\chi}^{(2)}\Big[ V_{SSP}^{(+)}  \int_{\rm{ring}} \mathrm{d}  \boldsymbol{r}_\perp W_{SSP}^{(+)}(\boldsymbol{r}_\perp) \\
    & \,\,\,\,\,\,\,\,\,\,\,\,\,\,\,\,\,\,\,\,\,\,\,\,\,\, + V_{SSP}^{(-)} \int_{\rm{ring}} \mathrm{d} \boldsymbol{r}_\perp W_{SSP}^{(-)}(\boldsymbol{r}_\perp)\Big].
\end{split}
    \label{eq:K_bar_RingFrameDegen}
\end{equation}

\subsection{Continuous-wave excitation}
For cw excitation, we find the generation rate similar to the calculation in Sec. \ref{subsec:cw}. We define 
\begin{equation}
    \hat{\mathsf{M}}_{\lambda\lambda'} (k_1,k_2) \equiv \sqrt{\frac{2 \pi P_P}{\hbar \omega_0 v_P^{\mathrm{ac}}}}  \hat{\mathsf{K}}_{\lambda\lambda'} (k_1,k_2,k_0), 
    \label{eq:MDefinitionDegen}
\end{equation}
and from a Fermi's golden rule calculation (\cite{bib:Banic2022}), we obtain
\begin{equation}
    \hat{R}^{SS} = \sum_{\lambda\lambda'}\frac{4 \pi}{\hbar^2} \int \mathrm{d} k_1 \mathrm{d} k_2 \delta(\Omega_{SS}(k_1,k_2)) |\hat{\mathsf{M}}_{\lambda\lambda'} (k_1,k_2)|^2.
    \label{eq:Rate_PairsDegen}
\end{equation}
The generation rate of photons pairs within the same frequency bin $S$, for which one leaves by the $\lambda$ waveguide and the other through the $\lambda'$ waveguide is
\begin{equation}
\begin{split}
    \hat{R}^{SS}_{\lambda\lambda'} = & \frac{ P_P \hat{P}_{\rm{vac}} \left(\bar{\chi}^{(2)}\right)^2 \sqrt{\omega_S\omega_S} \bar{\Gamma}_P \eta_{S}^{\lambda} \eta_{S}^{\lambda'} \eta_P^{\rm{(ac)}} }{4 \hbar \pi \epsilon_0 c^3 \bar{\Gamma}_S \bar{\Gamma}_S \mathcal{R}} \\
    & \,\,\,\,\,\,\,\,\,\,\,\,\,\,\,\,\,\,\,\, \times {\frac{\bar{v}_{S} \bar{v}_{S} \bar{v}_P}{\bar{n}_{S} \bar{n}_{S} \bar{n}_P}} \frac{1}{A_{\mathrm{eff}}^{(\pm)}} \frac{1}{{\left[(\delta\omega)^2+(\bar{\Gamma}_P)^2\right]}},
\end{split} \label{eq:Rate_Pairs_Total_ResonanceQuality_QuasiPhaseMatchedDegen}
\end{equation}
where $\hat{P}_\mathrm{vac}$ is given by
\begin{equation}
    \hat{P}_{\rm{vac}} = \frac{\hbar}{2}\frac{\sqrt{\omega_S\omega_S}\bar{\Gamma}_S \bar{\Gamma}_S (\bar{\Gamma}_S + \bar{\Gamma}_S)}{(\Delta \omega)^2 + (\bar{\Gamma}_S + \bar{\Gamma}_S)^2}.
    \label{eq:VacPowerDegen}
\end{equation}
The generation efficiencies are given by
\begin{equation}
    \hat{\mathcal{E}}^{SS} = \sum_{\lambda\lambda'}\hat{\mathcal{E}}^{SS}_{\lambda\lambda'} = \sum_{\lambda\lambda'}\hat{{R}}^{SS}_{\lambda\lambda'} \frac{\hbar\omega_0}{P_P}, 
    \label{eq:GenEffHat}
\end{equation}
and the rate and efficiencies for different combinations of exit waveguides $\mu$ and $\mu'$ are given by
\begin{equation}
    \frac{\hat{R}^{SS}_{\lambda\lambda'}}{\hat{R}^{SS}_{\mu\mu'}} = \frac{\hat{\mathcal{E}}^{SS}_{\lambda\lambda'}}{\hat{\mathcal{E}}^{SS}_{\mu\mu'}} = \frac{\eta_{S}^{\lambda} \eta_{S}^{\lambda'}}{\eta_{S}^{\mu} \eta_{S}^{\mu'}} = \frac{\Gamma_S^{\lambda} \Gamma_S^{\lambda'}}{\Gamma_S^{\mu} \Gamma_S^{\mu'}}.
    \label{eq:RatesDifferentExitsDegen}
\end{equation}

\subsection{Pulsed excitation}
\subsubsection{{pair regime}}
In the {pair regime} using the same approach from Sec. \ref{subsub:PairRegime}, we write the generated ket (similar to \cite{bib:Drago}):
\begin{equation}
\begin{split}
    \ket{\hat{\psi}} &\approx \ket{\mathrm{vac}}+\frac{\hat{\beta}}{\sqrt{2}}\sum_{\lambda\lambda'}\ket{\mathrm{\hat{II}}}_{\lambda\lambda'} ,
    \end{split}
    \label{eq:OutputKetDrago}
\end{equation}
where $|\hat{\beta}|^2/2$ is the probability of generating a pair of photons. We write the normalized two-photon ket (i.e., $\braket{\hat{\mathrm{II}}}{\hat{\mathrm{II}}}=1$)
\begin{equation}
\ket{\hat{\mathrm{II}}} = \sum_{\lambda\lambda'}\ket{\mathrm{\hat{II}}}_{\lambda\lambda'} = (\hat{A}_{\mathrm{II}}^{\mathrm{out}})^\dagger\ket{\mathrm{vac}},
    \label{eq:Ket}
\end{equation}
where
\begin{equation}
\begin{split}
    (\hat{A}_{\mathrm{II}}^{\mathrm{out}})^\dagger \equiv \frac{1}{\sqrt{2}}\sum_{\lambda\lambda'}  & \int \mathrm{d}k_1 \mathrm{d}k_2 \hat{\varphi}^\mathrm{II}_{\lambda\lambda'}(k_1,k_2)\\
    &\times\left[{a}_{S}^{\mathrm{out},\lambda}(k_1)\right]^\dagger \left[ {a}_{S}^{\mathrm{out},\lambda'}(k_2)\right]^\dagger,
    \end{split}
    \label{eq:SuperADeg}
\end{equation}
and where the BWF is given by
\begin{equation}
\begin{split}
    \hat{\varphi}^{\mathrm{II}}_{\lambda\lambda'}(k_1,k_2) \equiv \frac{4\pi i \alpha}{\hat{\beta}v_P^{\mathrm{ac}}\hbar}&   \hat{\mathsf{K}}_{\lambda\lambda'}(k_1,k_2,X_P) \phi(X_P).
\end{split}
    \label{eq:BiphotonWFDeg}
\end{equation}
The BWF is normalized according to
\begin{equation}
\sum_{\lambda\lambda'}\int\mathrm{d}k_1\mathrm{d}k_2 \left|\hat{\varphi}^{\mathrm{II}}_{\lambda\lambda'}(k_1,k_2)\right|^2 = 1,
    \label{eq:NormBiphoton2}
\end{equation}
{where here, since we are considering degenerate SPDC, we have $\hat{\varphi}^{\mathrm{II}}_{\lambda\lambda'}(k_1,k_2)=\hat{\varphi}^{\mathrm{II}}_{{\lambda'\lambda}}(k_2,k_1)$.} By normalizing the BWF, we find the number of generated photon pairs exiting the waveguides $\lambda$ and $\lambda'$
\begin{equation}
\begin{split}
    \frac{|\hat{\beta}_{\lambda\lambda'}|^2}{2} =& \frac{8\pi^2 E_P}{(v_P^{\mathrm{ac}})^2\hbar^3\omega_0}\\
    & \times \int\mathrm{d}k_1\mathrm{d}k_2 \left|  \hat{\mathsf{K}}_{\lambda\lambda'}(k_1,k_2,X_P)   \phi \left(X_P\right)\right|^2,
    \label{eq:betaHat}
\end{split}
\end{equation}
where the total probability of generating a pair is given by
\begin{equation}
    \frac{|\hat{\beta}|^2}{2} = \sum_{\lambda\lambda'} \frac{|\hat{\beta}_{\lambda\lambda'}|^2}{2}.
    \label{eq:SumBetaHat}
\end{equation}
The {pair regime} efficiency is given by
\begin{equation}
\begin{split}
    \hat{\mathcal{E}}_\mathrm{pulse; pair regime}^{SS} & = \frac{|\hat{\beta}|^2}{2|\alpha|^2},
    \label{eq:LowGainHat}
    \end{split}
\end{equation}
and the efficiency for each output channel is
\begin{equation}
\begin{split}
    [\hat{\mathcal{E}}_\mathrm{pulse; pair regime}^{SS}]_{\lambda\lambda'} & = \frac{|\beta_{\lambda\lambda'}|^2}{2|\alpha|^2},
    \label{eq:LowGainEachHat}
    \end{split}
\end{equation}
and similarly to Eq. \eqref{eq:RatesDifferentExitsDegen}, we have 
\begin{equation}
\begin{split}
    \frac{|\hat{\beta}_{\lambda\lambda'}|^2}{|\hat{\beta}_{\mu\mu'}|^2} = \frac{[\hat{\mathcal{E}}_\mathrm{pulse; pair regime}^{SS}]_{\lambda\lambda'}}{[\hat{\mathcal{E}}_\mathrm{pulse; pair regime}^{SS}]_{\mu\mu'}} = \frac{\eta_{S}^{\lambda} \eta_{S}^{\lambda'}}{\eta_{S}^{\mu} \eta_{S}^{\mu'}} = \frac{\Gamma_S^{\lambda} \Gamma_S^{\lambda'}}{\Gamma_S^{\mu} \Gamma_S^{\mu'}}.
    \label{eq:FracHat}
\end{split}
\end{equation}

\subsubsection{{Lowest-order squeezing regime}}
For the {LOS regime}, using the same approach {applied} in Sec. \ref{subsub:HighGain}, the generated state is written as 
\begin{equation}
\begin{split}
    \ket{\hat{\psi}}=& e^{\left[\frac{\hat{\beta}}{\sqrt{2}}(\hat{A}_{\mathrm{II}}^{\mathrm{out}})^\dagger -\mathrm{H.c.}\right]}\ket{\mathrm{vac}},
    \end{split}
    \label{eq:OutKetFinal2Degen2}
\end{equation}
where {now} $\hat{\beta}$ is the squeezing parameter \cite{bib:Drago}, and the degenerate pair generation operator $\hat{A}_{\mathrm{II}}^{\mathrm{out}}$ is as defined above. We can discretize the pair generation operator similarly to Eq. \eqref{eq:SuperAOutDiscrete}
\begin{equation}
    \begin{split}
\left(\hat{A}_\mathrm{II}^{\mathrm{out}}\right)^\dagger = \frac{1}{\sqrt{2}} & \sum_{\lambda\lambda'} \sum_{ij} \sqrt{\Delta k_1 \Delta k_2}\\
& \times \hat{\Phi}^{\mathrm{II}(ij)}_{\lambda\lambda'}\left[{a}_{Si}^{\mathrm{out},\lambda} \right]^\dagger \left[{a}_{Sj}^{\mathrm{out},\lambda'}\right]^\dagger.  
    \end{split}
    \label{eq:SuperAOutDiscreteHat}
\end{equation}
We combine the indices $\lambda$ and $i$ to one index $m$, and the indices $\lambda'$ and $j$ to $m'$. We can write
\begin{equation}
\begin{split}
    \left( \hat{A}_\mathrm{II}^{\mathrm{out}}\right)^\dagger = \frac{1}{\sqrt{2}} \sum_{mm'} \sqrt{\Delta k_1 \Delta k_2} \hat{\Phi}^{\mathrm{II}}_{mm'}\left[{a}_{S}^{\mathrm{out},m}\right]^\dagger \left[ {a}_{S}^{\mathrm{out},m'}\right]^\dagger.
    \end{split}
    \label{eq:SuperAOutDiscrete2Hat}
\end{equation}
We can define the symmetric squeezing matrix $\hat{\boldsymbol{J}}$ \cite{Vendromin24}, with elements  
\begin{equation}
\begin{split}
    \hat{J}_{mm'} = \hat{\beta} \sqrt{\Delta k_1 \Delta k_2} \hat{\Phi}^{\mathrm{II}}_{mm'},
    \end{split}
    \label{eq:JMatrixHat}
\end{equation}
and we write
\begin{equation}
\begin{split}
    \left( \hat{A}_\mathrm{II}^{\mathrm{out}}\right)^\dagger = \frac{1}{\hat{\beta}\sqrt{2}} \sum_{mm'}  \hat{J}_{mm'} \left[{a}_{S}^{\mathrm{out},m} \right]^\dagger \left[{a}_{S}^{\mathrm{out},m'}\right]^\dagger,
    \end{split}
    \label{eq:SuperAOutDiscrete3Hat}
\end{equation}
allowing us to write the degenerate generated ket 
\begin{equation}
\begin{split}
    \ket{\hat{\psi}} =& \exp \left\{\sum_{mm'} \frac{1}{2} \hat{J}_{mm'} \left[{a}_{S}^{\mathrm{out},m}\right]^\dagger \left[ {a}_{S}^{\mathrm{out},m'}\right]^\dagger -\mathrm{H.c.}\right\}\ket{\mathrm{vac}}.
    \end{split}
    \label{eq:OutKetFinal4Hat}
\end{equation}
The moments for the degenerate squeezed state in Eq. \eqref{eq:OutKetFinal4Hat} have been calculated previously by Quesada et al \cite{bib:Quesada2022}. To calculate the moments, a similar method is used to the one in sub Sec. \ref{subsec:Pulsed}, except now one does a Takagi factorization \cite{houde2024matrixdecompositionsquantumoptics} of the symmetric squeezing matrix.

\end{document}